\documentclass[11pt]{article}
\usepackage{graphicx}
\usepackage{color}
\usepackage{amsmath, amssymb, amsthm, ascmac, caption, comment, enumerate, ascmac, bm}
\usepackage{latexsym, mathabx, mathrsfs, mathtools, psfrag, setspace, listings}
\usepackage{newtxtext}
\usepackage{natbib}
\usepackage[stable]{footmisc}
\usepackage{multirow}

\mathtoolsset{showonlyrefs=true}

\definecolor{mycolor}{cmyk}{0.82, 0.23, 0.26, 0}
\usepackage[colorlinks=true, linkcolor=mycolor, filecolor=mycolor, urlcolor=mycolor, citecolor=mycolor, driverfallback=dvipdfmx]{hyperref}
\usepackage[margin = 2.3cm]{geometry}

\usepackage{titlesec}
\titleformat*{\section}{\Large\bfseries\sffamily}
\titleformat*{\subsection}{\large\bfseries\sffamily}
\titleformat*{\subsubsection}{\large\bfseries\sffamily}

\DeclareMathOperator*{\argmin}{argmin}

\DeclareMathOperator*{\Cov}{\mathrm{Cov}}
\DeclareMathOperator*{\Var}{\mathrm{Var}}
\DeclareMathOperator*{\sumi}{\sum_{\it i \in \mathcal{I}}}
\DeclareMathOperator*{\sumj}{\sum_{\it j \in \mathcal{I}}}

\renewcommand{\hat}{\widehat}
\renewcommand{\tilde}{\widetilde}

\renewcommand{\bar}{\overline}
\newcommand{\bbE}{\mathbb{E}}

\newcommand{\bbR}{\mathbb{R}}
\newcommand{\bbL}{\mathbb{L}}

\numberwithin{equation}{section}
\numberwithin{table}{section}
\numberwithin{figure}{section}
\def \baselinestretch{1.2} 
\allowdisplaybreaks[2]

\theoremstyle{definition}
\newtheorem{theorem}{Theorem}[section]
\newtheorem{assumption}{Assumption}[section]

\newtheorem{proposition}{Proposition}[section]
\newtheorem{remark}{Remark}[section]

\title{Causal Interpretation of Linear Social Interaction Models with Endogenous Networks}

\author{Tadao Hoshino\thanks{School of Political Science and Economics, Waseda University, 1-6-1 Nishi-waseda, Shinjuku-ku, Tokyo 169-8050, Japan. Email: \href{mailto:thoshino@waseda.jp}{thoshino@waseda.jp}.}}

\begin{document}
\maketitle

\begin{abstract}
This study investigates the causal interpretation of linear social interaction models in the presence of endogeneity in network formation under a heterogeneous treatment effects framework.
We consider an experimental setting in which individuals are randomly assigned to treatments while no interventions are made for the network structure.
We show that running a linear regression ignoring network endogeneity is not problematic for estimating the average direct treatment effect. However, it leads to sample selection bias and negative-weights problem for the estimation of the average spillover effect.
To overcome these problems, we propose using potential peer treatment as an instrumental variable (IV), which is automatically a valid IV for actual spillover exposure.
Using this IV, we examine two IV-based estimands and demonstrate that they have a local average treatment-effect-type causal interpretation for the spillover effect.
\end{abstract}

\newpage

\section{Introduction}

Many studies have estimated treatment spillover effects using an experimental approach that randomly assigns individuals to treatments to exploit the exogeneity of their peers' treatment status (e.g., \citealp{oster2012determinants, cai2015social, paluck2016changing, angelucci2019incentives}).
In most of these studies, although the treatment assignments are randomized, they often ignore the endogenous nature of existing networks, which can hinder the correct identification of the causal effects of treatment spillovers.

For example, \cite{paluck2016changing} investigated the impact of an anti-conflict intervention program on adolescents' attitudes and their propagation through friendship networks.
In friendship networks, non-combative students are more likely to be friends with students of the same mindset (i.e., homophily).
Therefore, even if we can observe a group of students engaging in conflict-mitigation activities together, determining whether this reflects the causal spillover effect from the program's participant friends or a mere correlation between their personalities (or both) is not straightforward.
In this example, although the treatments were completely randomized, whose treatment status matters to whom remained endogenously determined.\footnote{
    In other words, actual treatment exposure consists of a product of exogenous and endogenous factors.
    \cite{borusyak2020non} studied this situation in a more general framework than ours.
    }

One possible approach to circumvent the network endogeneity problem is to randomly assign peers, as in \cite{sacerdote2001peer}, \cite{zimmerman2003peer}, \cite{guryan2009peer}, and \cite{booij2017ability}.
A clear limitation of this approach is that it is not feasible in most situations.
Meanwhile, several studies consider specific regression models of social interactions to deal with the endogeneity of networks (e.g., \citealp{goldsmith2013social, hsieh2016social, johnsson2021estimation, jochmans2022peer}).
Their results are not directly applicable to the potential outcome framework with heterogeneous treatment effects, and simply extending them to a nonparametric causal model is often challenging due to the curse of dimensionality.
Therefore, considering empirical feasibility, it would be preferable to focus on a simple regression approach based on existing networks.
Although the causal interpretation of linear regression models under network interference has been studied, for example, in \cite{baird2018optimal}, \cite{DITRAGLIA20231589}, and \cite{vazquez2022identification,vazquez2023causal}, none of these accounts for network endogeneity.
The purpose of our study is to bridge this gap.
 
This paper is organized as follows.
In Section \ref{sec:pair}, to clarify the problem and find possible remedies, we consider a toy model in which treatments are randomly assigned and each unit has exactly one potential interacting partner.
We assume that whether one's outcome is influenced by his partner or not is determined endogenously.
We show that when the network endogeneity is overlooked, although the ordinary least squares (OLS) estimand can correctly capture the average direct effect from one's own treatment, it is biased for the average spillover effect due to the correlation between potential outcomes and network connectivity.
To account for the endogeneity issue, we use an instrumental variable (IV) approach.
We employ the partner's treatment assignment as an IV, which is a valid IV for spillover exposure.
Based on this IV, we prove that a two-stage least squares (2SLS) regression has a local average treatment effect (LATE) type causal interpretation.
Furthermore, as a more efficient alternative to 2SLS, we propose a weighted least squares (WLS) method with the same LATE interpretation.

Section \ref{sec:gen} extends the discussion in Section \ref{sec:pair} to a general model that allows each unit to have multiple peers.
Although potential heterogeneity in the network structure complicates the analysis, we can confirm that essentially the same results as in the toy model hold.
We also show that in this general model, the OLS estimand suffers not only from the endogeneity bias but also from a negative-weights problem.  

In general, statistical inference for fully heterogenous treatment effect models under non-identical data structures is a challenging task.
In Section \ref{sec:inference}, we consider several empirically tractable alternatives to perform statistical inference, including a randomization test.
In Section \ref{sec:empiric}, we revisit the data in \cite{paluck2016changing} and demonstrate that the spillover effect indeed exists even after controlling for the network endogeneity.
In the supplementary appendix, we present proofs of the technical results and results of some Monte Carlo simulations.

\section{A pair-interaction model}\label{sec:pair}

Suppose there are two non-overlapping samples: the focal sample $\mathcal{I}$ and the partner sample $\mathcal{J}$.
Only the focal sample is used to estimate causal effects.
Each focal unit $i \in \mathcal{I}$ has exactly one potential partner $j(i) \in \mathcal{J}$, and $j(i)$ cannot be a partner of the other focal units.
Suppose that $j(i)$ does not have to always interact with $i$.
For each pair $(i, j(i))$, let $A_{j(i)}$ denote a dummy variable indicating whether $j(i)$'s treatment affects $i$'s outcome.
The outcome and treatment of interest are denoted by $Y \in \bbR$ and $D \in \{0,1\}$, respectively.
For example, this situation occurs when $(i, j(i))$ represents a couple, $A_{j(i)}$ indicates whether they are living together, and $D_{j(i)}$ indicates their partner's lifestyle, such as diet (e.g., vegetarian), with $Y$ being a health outcome.
We define the effective treatment spillover variable as $S_i \coloneqq A_{j(i)}D_{j(i)}$.

Let $Y_i(d,s)$ be the potential outcome when $D_i = d$ and $S_i = s$.
The observed outcome is written as 
\begin{align*}
    Y_i = \sum_{(d,s) \in \{0,1\}^2}\bm{1}\{D_i = d, S_i = s\}Y_i(d, s).
\end{align*}
To focus solely on the endogeneity issue caused by link connectivity $A_{j(i)}$, we rule out cases of self-selected treatments. 
Specifically, we consider the following Bernoulli-type experimental setting:
\begin{assumption}\label{as:rct_pair}
The treatments $\{D_i\}_{i \in \mathcal{I} \cup \mathcal{J}}$ are mutually independent such that $\Pr(D_i = d) = p_d^\mathcal{I}$ and $\Pr(D_j = d) = p_d^\mathcal{J}$.
In addition, $(D_i, D_{j(i)})$ are independent of $(Y_i(d,s), A_{j(i)})$ for all $(d,s) \in \{0,1\}^2$.
\end{assumption}
If we allow for the dependence among the treatment assignments, it causes a significant difficulty in interpreting our estimands.\footnote{When the treatment assignments are dependent, even the average direct effect from one's own treatment may involve an identification issue (cf. Theorem 2 in \cite{vazquez2022identification}).}
For this reason, studies in the literature of network interference often focus on the Bernoulli design (e.g., \citealp{hu2022average}; \citealp{li2022random}).
As we will demonstrate later, even in this ideal experimental setting, the spillover effects cannot be correctly identified in the presence of network endogeneity (not to mention in other experimental designs).

We also assume that the data are independent and identically distributed (IID).
\begin{assumption}\label{as:iid_pair}
The potential outcomes $\{Y_i(d,s)\}_{(d,s) \in \{0,1\}^2}$ and link connections $A_{j(i)}$ are IID across $i \in \mathcal{I}$ such that $\Pr(A_{j(i)}= a) = p_a^A$.
\end{assumption}
We allow $A_{j(i)}$ to correlate with $Y_i(d,s)$, which is the source of endogeneity of concern.
In the example above, it is natural to imagine that $A_{j(i)}=1$ is more likely if $i$ and $j(i)$ share similar lifestyle preferences, which suggests that $A_{j(i)}$ and $Y_i(d,s)$ are dependent.

\subsection{Selection bias in the OLS estimation}

We first discuss the selection bias in the OLS regression.
Suppose that a researcher estimates a linear regression model of $Y_i$ on $(D_i, S_i)$ using the OLS estimator.
Define
\begin{align}\label{eq:pair_ls_def}
        (\beta_0^{ols}, \beta_d^{ols}, \beta_s^{ols}) = \argmin_{b_0, b_d, b_s} \bbE[(Y_i - b_0 - b_d D_i - b_s S_i)^2].
\end{align}
This characterization implicitly relies on the IID assumption.
Note that since $D_i$ and $S_i$ are independent, we do not consider including a cross-term between them.

Here, we define the "direct" and "spillover" treatment effects as follows:
\begin{align*}
    & \delta_i(s) \coloneqq Y_i(1, s) - Y_i(0, s), \qquad \bar \delta_i\coloneqq \delta_i(1) p_1^\mathcal{J} + \delta_i(0) p_0^\mathcal{J} \\
    & \tau_i(d) \coloneqq Y_i(d, 1) - Y_i(d, 0), \qquad \bar \tau_i\coloneqq \tau_i(1) p_1^\mathcal{I} + \tau_i(0) p_0^\mathcal{I}.
\end{align*}
We summarize the causal interpretations of $(\beta_d^{ols}, \beta_s^{ols})$ as follows:

\begin{proposition}\label{prop:pair_ls}
    Suppose that $(\beta_d^{ols}, \beta_s^{ols})$ is determined by \eqref{eq:pair_ls_def}.
    Then, under Assumptions \ref{as:rct_pair} and \ref{as:iid_pair}, we have
    \begin{align*}
        \text{(i)}
        & \;\; \beta_d^{ols} = \bbE[\bar \delta_i\mid A_{j(i)}= 1] p_1^A + \bbE[\delta_i(0) \mid A_{j(i)}= 0] p_0^A \\
        \text{(ii)}
        & \;\; \beta_s^{ols} = \bbE[\bar \tau_i\mid A_{j(i)}= 1] + \bar \eta_i \frac{p_0^A}{1 - p_1^A p_1^\mathcal{J}},
    \end{align*}
    where $\bar \eta_i \coloneqq \eta_i(1) p_1^\mathcal{I} + \eta_i(0) p_0^\mathcal{I}$, and $\eta_i(d) \coloneqq \bbE[Y_i(d,0) \mid A_{j(i)}= 1] - \bbE[Y_i(d,0) \mid A_{j(i)}= 0 ]$.
\end{proposition}

As shown in result (i), $\beta_d^{ols}$ is the weighted average of the average direct treatment effects for $A_{j(i)}= 1$ and $A_{j(i)}= 0$, which has a clear causal interpretation.
Hence, ignoring the endogeneity in $A_{j(i)}$ is harmless for identifying the average direct effect.
By contrast, result (ii) shows that the spillover effect parameter $\beta_s^{ols}$ is a summation of the average spillover effect for $A_{j(i)}= 1$ and the selection bias term.
This bias term originates from the correlation between $Y_i(d,0)$ and $A_{j(i)}$.

\subsection{Causal interpretation of the 2SLS and WLS estimation}

To circumvent the selection bias in the OLS, we use the IV method.
For the choice of the IV for $S_i$, one obvious candidate is $D_{j(i)}$, which is completely exogenous according to the experimental design and is a determinant of $S_i$.
More importantly, in practical terms, it is not necessary to search for other IV candidates.
Then, based on this IV, we investigate the causal interpretation of the 2SLS estimand $(\beta_0^{2sls}, \beta_d^{2sls}, \beta_s^{2sls})$, which we define as
\begin{align}\label{eq:pair_2sls_def}
        (\beta_0^{2sls}, \beta_d^{2sls}, \beta_s^{2sls}) = \argmin_{b_0, b_d, b_s} \bbE[(Y_i - b_0 - b_d D_i - b_s \bbL(S_i \mid D_i, D_{j(i)}))^2],
\end{align}
where $\bbL(S_i \mid D_i, D_{j(i)}) = \gamma_0 + \gamma_d D_i + \gamma_s D_{j(i)}$ is the linear projection of $S_i$ onto $(D_i, D_{j(i)})$ obtained from the first-stage regression $(\gamma_0, \gamma_d, \gamma_s) = \argmin_{a_0, a_d, a_s} \bbE[(S_i - a_0 - a_d D_i - a_s D_{j(i)})^2]$.
Because $D_i$ and $S_i$ are independent, instrumenting for $S_i$ does not alter the interpretation of $\beta_d^{2sls}$; that is, $\beta_d^{2sls} = \beta_d^{ols}$ holds.
For a causal interpretation of $\beta_s^{2sls}$, we obtain the following result:

\begin{proposition}\label{prop:pair_2sls}
    Suppose that $\beta_s^{2sls}$ is determined by \eqref{eq:pair_2sls_def}.
    Then, under Assumptions \ref{as:rct_pair} and \ref{as:iid_pair}, we have $\beta_s^{2sls} =  \bbE[\bar \tau_i\mid A_{j(i)}= 1]$.
\end{proposition}

\begin{remark}[LATE interpretation 1]
In terms of the relationship between $S_i$ and $D_{j(i)}$, the ``complier'' group is defined in the usual sense as the set of individuals who receive $S_i = 1$ only when $D_{j(i)} = 1$.
Clearly, $i$ is a complier if and only if $A_{j(i)}= 1$.
Thus, as in the standard 2SLS estimation, Proposition \ref{prop:pair_2sls} shows that $\beta_s^{2sls}$ can be interpreted as the local average treatment effect (LATE) for the compliers in our context.
\end{remark}

\begin{remark}[LATE interpretation 2]
Using simple algebra, we obtain the following:
\begin{align}\label{eq:pair_2sls_exp}
    \beta_s^{2sls} 
    & = \frac{\bbE[(D_{j(i)} - \bbL(D_{j(i)} \mid D_i)) Y_i]}{\bbE[(D_{j(i)} - \bbL(D_{j(i)} \mid D_i)) S_i]},
\end{align}
where $\bbL(D_{j(i)} \mid D_i)$ denotes the linear projection of $D_{j(i)}$ onto $D_i$.
According to Proposition 1 in \cite{blandhol2022tsls}, for $\beta_s^{2sls}$ to have a LATE interpretation, $\bbL(D_{j(i)} \mid D_i) = \bbE[D_{j(i)} \mid D_i]$ is needed.
This equality holds trivially in our setting, with $\bbE[D_{j(i)} \mid D_i] = p_1^\mathcal{J}$.
Then, Proposition \ref{prop:pair_2sls} is a special case of their result.
\end{remark}

It is known that we can improve the estimation efficiency of the LATE parameter by weighting each observation according to its compliance probability (e.g., \citealp{joffe2003weighting, coussens2021improving}).
The same discussion applies here.
Moreover, because we can precisely identify each unit's compliance status (i.e., $A_{j(i)}$), the resulting estimator is reduced to a simple least squares regression of $Y_i$ on $(D_i, D_{j(i)})$ for those satisfying $A_{j(i)} = 1$.
Then, we define the following weighted least squares (WLS) estimand:
\begin{align}\label{eq:pair_sls_def}
        (\beta_0^{wls}, \beta_d^{wls}, \beta_s^{wls}) = \argmin_{b_0, b_d, b_s} \bbE[A_{j(i)} (Y_i - b_0 - b_d D_i - b_s D_{j(i)})^2].
\end{align}
Note that if every $i$ is affected by his/her partner, then the OLS and WLS coincide.
The next proposition provides a causal interpretation of the WLS estimand.

\begin{proposition}\label{prop:pair_sel_ls}
    Suppose that $(\beta_d^{wls}, \beta_s^{wls})$ is determined by \eqref{eq:pair_sls_def}.
    Then, under Assumptions \ref{as:rct_pair} and \ref{as:iid_pair}, we have (i) $\beta_d^{wls} = \bbE[\bar \delta_i\mid A_{j(i)}= 1]$ and (ii) $\beta_s^{wls} =  \bbE[\bar \tau_i\mid A_{j(i)}= 1]$.
\end{proposition}

Because of the sample selection, the interpretation of the direct effect $\beta_d^{wls}$ is slightly different from that for $\beta_d^{ols}$ and $\beta_d^{2sls}$.
The average direct effect for the non-compliers $\bbE[\delta_i(0) \mid A_{j(i)}= 0]$ is not incorporated in $\beta_d^{wls}$.
Meanwhile, the interpretation of $\beta_s^{wls}$ is the same as that of $\beta_s^{2sls}$. 

\paragraph{Numerical example}
Here, we briefly demonstrate the severity of the selection bias.
The simulation setup is as follows: $D_i, D_{j(i)} \sim \text{Bernoulli}(0.5)$, $A_{j(i)} = \bm{1}\{e^{(U_i + U_{j(i)})}/[1 + e^{(U_i + U_{j(i)})}] > 0.5\}$, where $U_i, U_{j(i)} \sim \text{Uniform}(-1,1)$, and $Y_i = \xi_i + U_i$, where $\xi_i \sim N(1,1)$.
The sample size is $n = 1000$.
By construction, both the direct treatment effect and spillover effect do not exist.
The results of OLS regression, 2SLS, and WLS are summarized in the next table.

\begin{table}[ht]
    \caption{A numerical simulation}
    \begin{center}
\begin{tabular}{l|cccccc}
    \hline\hline
     & \multicolumn{2}{c}{OLS} & \multicolumn{2}{c}{2SLS} & \multicolumn{2}{c}{WLS} \\
    Variable & Coef. & t-value & Coef. & t-value & Coef. & t-value \\
    \hline
    $D$ & 0.048 & 0.656 & 0.047 & 0.638 & 0.097 & 0.973\\
    $S$ & 0.553 & 6.666 & 0.136 & 0.924 & 0.046 & 0.463\\
    \hline
\end{tabular}
\end{center}
\end{table}

The table shows that in the presence of network endogeneity, the simple OLS regression incorrectly detects the spillover effect, even though the treatments are completely randomly assigned.
In contrast, the 2SLS and WLS estimators correctly evaluate the spillover effect.
The replication R code for this experiment is provided in Appendix \ref{app:Rcode}.

\paragraph{Efficiency comparison}

In Appendix \ref{app:proof_sec2}, under standard moment conditions, we show that the sample analog of the 2SLS estimand and that of WLS are both asymptotically normal in the following sense:
\begin{align*}
    \sqrt{n}(\hat \beta_s^{2sls} - \beta_s^{2sls})
    \overset{d}{\to} N \left(0, \frac{p_0^\mathcal{J} \sigma^2_{\varepsilon}(1) + p_1^\mathcal{J} \sigma^2_{\varepsilon}(0) }{(p_1^A)^2 p_1^\mathcal{J} p_0^\mathcal{J}}\right),
    \qquad 
    \sqrt{n}(\hat \beta_s^{wls} - \beta_s^{2sls})
    \overset{d}{\to} N \left(0, \frac{p_0^\mathcal{J} \sigma^2_{\epsilon,1}(1) + p_1^\mathcal{J} \sigma^2_{\epsilon,1}(0) }{p_1^A p_1^\mathcal{J} p_0^\mathcal{J}}\right)
\end{align*}
where $n = |\mathcal{I}|$, $\sigma^2_{\varepsilon}(d) \coloneqq \bbE[\varepsilon_i^2 \mid D_{j(i)} = d]$, $\varepsilon_i \coloneqq Y_i - \beta_0^{2sls} - \beta_d^{2sls} D_i - \beta_s^{2sls} S_i$, $\sigma^2_{\epsilon, 1}(d) \coloneqq \bbE[\epsilon_{1, i}^2 \mid D_{j(i)} = d, A_{j(i)} = 1]$, and $\epsilon_{1, i} \coloneqq Y_i - \beta_0^{wls} - \beta_d^{wls} D_i - \beta_s^{wls} D_{j(i)}$.
This is a common result for the 2SLS estimator: the asymptotic variance is inversely proportional to the ``square'' of the compliance probability.
On the other hand, the asymptotic variance of the WLS estimator is inversely proportional only to $p_1^A$.
In particular, for the case of homoscedasticity such that $\sigma^2_{\varepsilon}(d) = \sigma^2_{\epsilon, 1}(d)$, the WLS estimator is more efficient than the 2SLS estimator exactly  by a factor of $p_1^A$.

\section{Linear social interaction models with a general network}\label{sec:gen}

In this section, we generalize the above discussion to models in which interactions can occur among more than two individuals.
For each $i \in \mathcal{I}$, let $\mathcal{P}_i \subseteq \mathcal{J}$ be a group of potential peers, such as family members, classmates, and local neighborhoods, depending on the context.
The size of $\mathcal{P}_i$ is denoted by $n_i \coloneqq |\mathcal{P}_i|$ and may vary across $i$ and $n_i \ge 1$ for all $i \in \mathcal{I}$.
We assume that $\mathcal{P}_i$ and $\mathcal{P}_{i'}$ are disjoint for any $i \neq i'$.

It is important to note that in the following analysis, $\mathcal{P}_i$ is treated as fixed for each $i$ (which may or may not be endogenous in reality).
Thus, the subsequent discussion should be viewed as being conditional on $\mathcal{P}_i$'s.
If we allow them to be random, then we will need to take into account the correlation between the potential outcomes and the composition and size of $\mathcal{P}_i$, which significantly complicates the analysis.\footnote{
    Insofar as the analysis is based on existing networks, this level of network endogeneity may not be addressed without additional information, such as the availability of individual covariates to instrument $\mathcal{P}_i$.
    For example, assuming that $\mathcal{P}_i$ is exogenously constructed once conditioned on each school, then we can consider segmenting the data by school to control for the endogeneity.
    Otherwise, if we want to obtain ``unconditional'' spillover effects, we might need to resort to an experimental approach that intervenes in the network structure itself.
}

Denoting the elements of $\mathcal{P}_i$ as $\mathcal{P}_i = \{1(i), \ldots, n_i(i)\}$, the link connections are characterized by $\bm{A}_{\mathcal{P}_i} = (A_{1(i)}, \ldots, A_{n_i(i)})^\top$, whose support is $\mathcal{A}_i \coloneqq \{0, 1\}^{n_i}$, where $A_{j(i)} = 1$ means that the treatment of $j$-th peer affects $i$'s outcome.
The peer treatments are denoted by $\bm{D}_{\mathcal{P}_i} \coloneqq (D_{1(i)}, \ldots, D_{n_i(i)})^\top$.
Then, the number of treated effective peers can be written as $R_i \coloneqq \bm{A}_{\mathcal{P}_i}^\top \bm{D}_{\mathcal{P}_i}$, which ranges over $\mathcal{R}_i \coloneqq \{0,1, \ldots, n_i\}$.

We assume that $R_i$ contains sufficient information on the treatment spillover effects in the sense that 
\begin{align*}
    Y_i = \sum_{(d,r) \in \{0,1\} \times \mathcal{R}_i} \bm{1}\{D_i = d, R_i = r\} Y_i(d,r),
\end{align*}
where $Y_i(d,r)$ denotes the potential outcome when $D_i = d$ and $R_i = r$.
This implicitly imposes anonymity and homogeneity in the treatment spillover mechanism, which is a standard assumption in the literature.
It is worth noting that $R_i$ becomes an exogenous variable when conditioned on $i$'s degree: $\bar A_i \coloneqq \sum_{j = 1}^{n_i} A_{j(i)}$.
That is, we have $\bbE[Y_i \mid R_i = r, D_i = d, \bar A_i = \bar a] = \bbE[Y_i(d,r) \mid \bar A_i = \bar a]$, implying that we can identify the average direct and spillover effects conditional on $\bar A_i = \bar a$ by a nonparametric regression of $Y_i$ on $(R_i, D_i, \bar A_i)$ (cf. \citealp{leung2020treatment}).
However, since all these regressors are discrete, performing this nonparametric regression is impractical due to the curse of dimensionality.
Instead, in practice, researchers often employ a linear social interaction model as follows:
\begin{align*}
    Y_i = \beta_0 + \beta_d D_i + \beta_s M_i + \varepsilon_i,
\end{align*}
where $M_i: \mathcal{R}_i \to \mathbb{R}$ is a known non-decreasing transformation of $R_i$.
Common choices for $M_i$ would be $M_i(r) = r$, $M_i(r) = r/n_i$, and $M_i(r) = \bm{1}\{r > 0\}$.\footnote{
    These three $M_i$ transformations are deterministic.
    Another empirically common choice is $M_i(r) = r/\bar A_i$.
    Since $\bar A_i$ is random and may be correlated with the potential outcomes, this case needs to be discussed separately from the others.
    Fortunately, it turns out that our causal interpretations for the 2SLS and WLS estimands remain valid for this $M_i$ too -- see the proofs of Theorems \ref{thm:net_2sls} and \ref{thm:net_sls}. 
}
Note again that since $D_i$ and $M_i$ are independent, a cross-term between them is not included in the model.

To facilitate the analysis, we assume an experimental setup similar to that considered above.

\begin{assumption}\label{as:rct_net}
The treatments $\{D_i\}_{i \in \mathcal{I} \cup \mathcal{J}}$ are mutually independent such that $\Pr(D_i = d) = p_d^\mathcal{I}$ and $\Pr(D_j = d) = p_d^\mathcal{J}$.
In addition, $(D_i, \bm{D}_{\mathcal{P}_i})$ are independent of $(Y_i(d,r), \bm{A}_{\mathcal{P}_i})$ for all $(d,r) \in \{0,1\} \times \mathcal{R}_i$.
\end{assumption}
\begin{assumption}\label{as:id_net}
The potential outcomes $\{Y_i(d,r)\}_{(d,r) \in \{0,1\} \times \mathcal{R}_i}$ and link connections $\bm{A}_{\mathcal{P}_i}$ are independent across $i \in \mathcal{I}$.
\end{assumption}

\subsection{Interpretation problems in the OLS estimation}

We first characterize the selection bias in the OLS estimand caused by network endogeneity.
The parameters of interest are as follows:
\begin{align}\label{eq:ls_net_def}
    (\beta_0^{ols}, \beta_d^{ols}, \beta_s^{ols}) = \argmin_{b_0, b_d, b_s} \sumi\bbE[(Y_i - b_0 - b_d D_i - b_s M_i)^2].
\end{align}
In contrast to the previous case, because the data may be non-identically distributed owing to the heterogeneity in $\mathcal{P}_i$, the target parameters essentially depend on the specific composition of $\mathcal{I}$ (but we suppress the dependence for notational simplicity).
Note that the estimator does not use the information of $\mathcal{P}_i$, and thus it can be unobserved to us as long as $R_i$ is known.
The same is true for the 2SLS and WLS.

Define
\begin{align*}
    \delta_i(r) \coloneqq Y_i(1, r) - Y_i(0, r), \qquad
    \tau_i^0(d, r) \coloneqq Y_i(d, r) - Y_i(d, 0), \qquad
    \bar \tau_i^0(r) \coloneqq \tau_i^0(1, r) p_1^\mathcal{I} + \tau_i^0(0, r) p_0^\mathcal{I}.
\end{align*}
Here, $\tau_i^0(d, r)$ measures the spillover effect, using $Y_i(d, 0)$ as the baseline.

\begin{theorem}\label{thm:net_ls}
Suppose that $(\beta_d^{ols}, \beta_s^{ols})$ is determined by \eqref{eq:ls_net_def}.
Then, under Assumptions \ref{as:rct_net} and \ref{as:id_net}, we have 
\begin{align*}
    \text{(i)} 
    & \;\; \beta_d^{ols} = \frac{1}{n}\sumi \sum_{(r, \vec{a}) \in \mathcal{R}_i \times \mathcal{A}_i} \bbE[\delta_i(r) \mid \bm{A}_{\mathcal{P}_i} = \vec{a}] \Pr(R_i = r, \bm{A}_{\mathcal{P}_i} = \vec{a}) \\
    \text{(ii)}
    & \;\; \beta_s^{ols} = \frac{1}{n} \sumi \sum_{(r, \vec{a}) \in \mathcal{R}_i \times \mathcal{A}_i} \pi_i(r, \vec{a}) \left\{ \bbE[\bar \tau_i^0(r) \mid \bm{A}_{\mathcal{P}_i} = \vec{a}] + \bbE[\bar Y_i(0) \mid \bm{A}_{\mathcal{P}_i} = \vec{a}] \right\},
\end{align*}
where $\pi_i(r, \vec{a}) \coloneqq \frac{(M_i(r) - \mu) \Pr(R_i = r, \bm{A}_{\mathcal{P}_i} = \vec{a})}{n^{-1}\sumi \bbE[(M_i - \mu)^2]}$, $\mu \coloneqq \frac{1}{n} \sumi \bbE[M_i]$, and $\bar Y_i(0) \coloneqq Y_i(1,0) p_1^\mathcal{I} + Y_i(0, 0) p_0^\mathcal{I}$.
\end{theorem}

From Theorem \ref{thm:net_ls}, $\beta_d^{ols}$ can be expressed as a weighted average of the conditional average direct effects.
Thus, it does not lose causal interpretability even if the network endogeneity is ignored, similar to Proposition \ref{prop:pair_ls}.
On the other hand, $\beta_s^{ols}$ includes the causal effect term $\sum_{(r, \vec{a}) \in \mathcal{R}_i \times \mathcal{A}_i} \pi_i(r, \vec{a}) \bbE[\bar \tau_i^0(r) \mid \bm{A}_{\mathcal{P}_i} = \vec{a}]$ and the selection bias term $\sum_{(r, \vec{a}) \in \mathcal{R}_i \times \mathcal{A}_i} \pi_i(r, \vec{a})\bbE[\bar Y_i(0) \mid \bm{A}_{\mathcal{P}_i} = \vec{a}]$.
The bias comes from the correlation between $Y_i(d,0)$ and $\bm{A}_{\mathcal{P}_i}$.
Moreover, even if the selection bias happens to be zero, the causal effect term is not purely causal because some weights $\{\pi_i(r, \vec{a})\}$ can be negative: indeed, we can easily see that $\frac{1}{n} \sumi \sum_{(r, \vec{a}) \in \mathcal{R}_i \times \mathcal{A}_i} \pi_i(r, \vec{a}) = 0$.
This is a tricky problem that does not arise in models with binary spillover exposure, which was originally pointed out by \cite{vazquez2022identification}.

\subsection{Causal interpretation of the 2SLS and WLS estimation}

To perform the 2SLS estimation, a natural IV candidate for $R_i$ would be the summation of the peers' treatments $\sum_{j = 1}^{n_i} D_{j(i)}$.
However, this choice of IV is not favorable in terms of obtaining a clear causal interpretation because $R_i$ is not directly a function of it.
Another possibility is to use $\bm{D}_{\mathcal{P}_i}$ as separate IVs.
With this approach, we could define various compliance types according to the connectivity of $A_{j(i)}$'s, such as $D_{1(i)}$-complier, $D_{2(i)}$-complier, etc, as in \cite{mogstad2021causal}.
Note however that even for a two-IV case, there are four different compliance patterns (see Table 2 in \cite{mogstad2021causal}).
More importantly, we allow each unit to have a heterogeneous number of peers.
This means that for some units, some $D_{j(i)}$'s may be non-existent, which further complicates the interpretation and raises practical feasibility concerns.
In addition, the peers are neither ordered nor labelled in general.

Considering these points, in the following, we focus on the case where a "single" element of $\bm{D}_{\mathcal{P}_i}$ is used as the IV for $R_i$.
Without loss of generality, suppose that the IV is the first element of $\bm{D}_{\mathcal{P}_i}$, $D_{1(i)}$.
Then, the potential treatment when $D_{1(i)} = d$ can be written as
\begin{align*}
    R_i(d) \coloneqq A_{1(i)} d + A_{2(i)} D_{2(i)} + \cdots + A_{n_i(i)} D_{n_i(i)}.
\end{align*}
The observed treatment is $R_i = D_{1(i)} R_i(1) + (1 - D_{1(i)}) R_i(0)$.
By definition, the monotonicity condition $R_i(1) \ge R_i(0)$ holds trivially, and the inequality is strict if and only if $A_{1(i)} = 1$.\footnote{
    As pointed out by a referee, as the size of potential peers $n_i$ increases, the contribution of each peer member decreases, which may lead to a weak IV problem.
    Therefore, when $n_i$ is relatively large, it might be better to reconsider the possibility of using multiple IVs.
    Also, the strength of $D_{1(i)}$'s as IVs is maximised when $A_{1(i)} =1$ for all $i$ (which is the source of the efficiency improvement in the WLS estimator).
    In this sense, since the choice of ``first peer'' is completely arbitrary, one should choose a person who is actually a friend of $i$ as $1(i)$.
}

With this IV, the population 2SLS estimand $(\beta_0^{2sls}, \beta_d^{2sls}, \beta_s^{2sls})$ is defined as
\begin{align}\label{eq:2sls_net_def}
    (\beta_0^{2sls}, \beta_d^{2sls}, \beta_s^{2sls}) = \argmin_{b_0, b_d, b_s} \sumi \bbE[(Y_i - b_0 - b_d D_i - b_s \bbL(M_i \mid D_i, D_{1(i)}))^2],
\end{align}
 where $\bbL(M_i \mid D_i, D_{1(i)}) \coloneqq \gamma_0 + \gamma_d D_i + \gamma_s D_{1(i)}$, and $(\gamma_0, \gamma_d, \gamma_s) = \argmin_{a_0, a_d, a_s} \sumi \bbE[(M_i - a_0 - a_d D_i - a_s D_{1(i)})^2]$.
As in the pair-interaction model, as $D_i$ is independent of $M_i$, $\beta_d^{ols} = \beta_d^{2sls}$ holds.
Thus, the same characterization as in Theorem \ref{thm:net_ls}(i) applies to $\beta_d^{2sls}$.

Let
\begin{align*}
    \tau_i^1 (d, r) \coloneqq Y_i(d, r) - Y_i(d, r - 1), 
    \qquad \bar \tau_i^1(r) \coloneqq \tau_i^1(1, r)p_1^\mathcal{I} + \tau_i^1(0, r)p_0^\mathcal{I}.
\end{align*}
The next theorem presents a causal interpretation of $\beta_s^{2sls}$.

\begin{theorem}\label{thm:net_2sls}
    Suppose that $\beta_s^{2sls}$ is determined by \eqref{eq:2sls_net_def}.
    Then, under Assumptions \ref{as:rct_net} and \ref{as:id_net}, we have 
\begin{align*}
    \beta_s^{2sls} = \frac{1}{n} \sumi \sum_{r = 1}^{n_i} \omega_i(r) \bbE[ \bar \tau_i^1(r) \mid R_i(1) \ge r > R_i(0)] ,
\end{align*}
where $\omega_i(r) \coloneqq \frac{\Pr( R_i(1) \ge r > R_i(0))}{n^{-1} \sumi \bbE[M_i^1 - M_i^0]}$ and $M_i^d \coloneqq M_i(R_i(d))$.
\end{theorem}

\begin{remark}[LATE interpretation 3]\label{rem:late3}
    Theorem \ref{thm:net_2sls} demonstrates that the 2SLS estimand $\beta_s^{2sls}$ has a causal interpretation as the weighted sum of the average treatment effects, where all weights $\{\omega_i(r)\}$ are positive.
    A special case where $\beta_s^{2sls}$ can be interpreted as a weighted "average" of the average treatment effects is when $M_i(\cdot)$ is the identity mapping; that is, $\bbE[M_i^1 - M_i^0] = \sum_{r = 1}^{n_i}\Pr(R_i(1) \ge r > R_i(0))$.
    Noting the equivalence of $\{R_i(1) \ge r > R_i(0)\}$ and $\{A_{1(i)} = 1, R_i(1) = r\}$, we can view $\beta_s^{2sls}$ as the weighted sum of LATEs: $\bbE[ \bar \tau_i^1(r) \mid A_{1(i)} = 1, R_i(1) = r]$, each corresponding to a subset of compliers sorted by $R_i(1)$.
\end{remark}

From the same argument as before, it is possible to improve the efficiency of the 2SLS estimator by appropriately weighting the data.
In this case, the compliers are those with a link connection with the first peer.
Thus, the WLS estimand $(\beta_0^{wls}, \beta_d^{wls}, \beta_s^{wls})$ is defined as
\begin{align}\label{eq:sls_net_def}
    (\beta_0^{wls}, \beta_d^{wls}, \beta_s^{wls}) = \argmin_{b_0, b_d, b_s} \sumi \bbE[A_{1(i)}(Y_i - b_0 - b_d D_i - b_s \mathbb{L}_1(M_i \mid D_i, D_{1(i)}))^2],
\end{align}
where $\mathbb{L}_1(M_i \mid D_i, D_{1(i)}) \coloneqq \gamma_{0,1} + \gamma_{d,1} D_i + \gamma_{s,1} D_{1(i)}$, and $(\gamma_{0,1}, \gamma_{d,1}, \gamma_{s,1}) = \argmin_{a_0, a_d, a_s} \sumi \bbE[A_{1(i)}(M_i - a_0 - a_d D_i - a_s D_{1(i)})^2]$.
Notice that if, for example, $1(i)$ represents $i$'s best friend and everyone has his/her best friend, the WLS is identical to the 2SLS.

\begin{theorem}\label{thm:net_sls}
    Suppose that $(\beta_d^{wls}, \beta_s^{wls})$ is determined by \eqref{eq:sls_net_def}.
    Then, under Assumptions \ref{as:rct_net} and \ref{as:id_net}, we have 
\begin{align*}
    \text{(i)} 
    & \;\; \beta_d^{wls} = \frac{n^{-1}\sumi \sum_{(r, \vec{a}) \in \mathcal{R}_i \times \{1, \mathcal{A}_{i,-1}\}} \bbE[\delta_i(r) \mid \bm{A}_{\mathcal{P}_i} = \vec{a}] \Pr(R_i = r, \bm{A}_{\mathcal{P}_i} = \vec{a})}{n^{-1} \sumi \Pr(A_{1(i)} = 1) } \\
    \text{(ii)}
    & \;\; \beta_s^{wls} = \frac{1}{n} \sumi \sum_{r = 1}^{n_i} \omega_i(r) \bbE[ \bar \tau_i^1(r) \mid R_i(1) \ge r > R_i(0)],
\end{align*}
where $\mathcal{A}_{i,-1} \coloneqq \{0,1\}^{n_i - 1}$.
\end{theorem}

As shown in Theorem \ref{thm:net_sls}(i), the causal interpretation of $\beta_d^{wls}$ is similar to that of $\beta_d^{ols}$ and $\beta_d^{2sls}$, except that it is conditioned on $A_{1(i)} = 1$.
Result (ii) shows that the WLS estimand $\beta_s^{wls}$ and the 2SLS estimand $\beta_s^{2sls}$ have a completely identical characterization; thus, the same LATE interpretation as in Remark \ref{rem:late3} applies.

\subsection{Comparison of the asymptotic distributions}\label{subsec:compare}

This subsection briefly discusses the asymptotic distributions of the 2SLS and WLS estimators.
The sample analog of $\beta_s^{2sls}$ can be obtained by $\hat \beta_s^{2sls} = \tilde{\bm{D}}_{1,n}^\top \bm{Y}_n/ \tilde{\bm{D}}_{1,n}^\top \bm{M}_n$, where $\bm{Y}_n = (Y_1, \ldots, Y_n)^\top$, $ \tilde{\bm{D}}_{1,n} \coloneqq \bm{D}_{1,n} - \bm{D}_{c,n}(\bm{D}_{c,n}^\top \bm{D}_{c,n})^{-1}\bm{D}_{c,n}^\top \bm{D}_{1,n}$, $D_{c,i} = (1, D_i)^\top$, $\bm{D}_{c,n} = (D_{c,1}, \ldots, D_{c,n})^\top$, $\bm{D}_{1,n} = (D_{1(1)}, \ldots, D_{1(n)})^\top$, and $\bm{M}_n = (M_1, \ldots, M_n)^\top$.
Furthermore, the population 2SLS residual can be written as $\varepsilon_i \coloneqq Y_i - \beta_0^{2sls} - \beta_d^{2sls} D_i - \beta_s^{2sls} M_i$.
Note that the population residuals may have unknown heterogeneous means caused by heterogeneity in the distribution of potential outcomes and network structure.
Hence, we introduce additional assumptions on the data to facilitate the derivation of the limiting distribution.
Specifically, suppose that $\bbE[\varepsilon_i] = \bbE[D_i \varepsilon_i] = \bbE[D_{1(i)} \varepsilon_i] = 0$.
Then,
\begin{align}\label{eq:clt_net_2sls}
    \sqrt{n}(\hat \beta_s^{2sls} - \beta_s^{2sls}) 
    & \overset{d}{\to} N\left(0, \lim_{n\to\infty}\frac{ \frac{1}{n} \sumi (p_0^\mathcal{J} \sigma^2_{\varepsilon,i}(1) + p_1^\mathcal{J} \sigma^2_{\varepsilon,i}(0))}{\left(\frac{1}{n}\sumi \bbE[M_i^1 - M_i^0] \right)^2 p_1^\mathcal{J} p_0^\mathcal{J}} \right),
\end{align}
where $\sigma^2_{\varepsilon,i}(d) \coloneqq \bbE[\varepsilon_i^2 \mid D_{1(i)} = d]$.
Noting the equality $\bbE[M_i^1 - M_i^0] = \bbE[M_i^1 - M_i^0 \mid A_{1(i)} = 1] \Pr(A_{1(i)} = 1)$, we can observe that the asymptotic variance is inversely related to the square of the compliance probability.

We now turn to the asymptotic distribution of the WLS estimator: $\hat \beta_s^{wls} = \tilde{\bm{D}}_{1,n,A}^\top \bm{Y}_n/ \tilde{\bm{D}}_{1,n,A}^\top \bm{M}_n$, where $\mathbb{I}_{n,A} \coloneqq \text{diag}(A_{1(1)}, \ldots, A_{1(n)})$, and $\tilde{\bm{D}}_{1,n,A} \coloneqq \mathbb{I}_{n,A} \bm{D}_{1,n} - \mathbb{I}_{n,A} \bm{D}_{c,n}(\bm{D}_{c,n}^\top \mathbb{I}_{n,A} \bm{D}_{c,n})^{-1} \bm{D}_{c,n}^\top \mathbb{I}_{n,A} \bm{D}_{1,n}$.
The population residual term is $\epsilon_i \coloneqq A_{1(i)} \epsilon_{1, i}$, where $\epsilon_{1, i} \coloneqq Y_i - \beta_0^{wls} - \beta_d^{wls} D_i - \beta_s^{wls} M_i$.
Similarly, we assume $\bbE[\epsilon_i] = \bbE[D_i \epsilon_i] = \bbE[D_{1(i)} \epsilon_i] = 0$.
Then, we have
\begin{align}\label{eq:clt_net_sls}
    \sqrt{n}(\hat \beta_s^{wls} - \beta_s^{wls})
    \overset{d}{\to} N \left(0, \lim_{n\to\infty} \frac{ \frac{1}{n}\sumi ( p_0^\mathcal{J} \sigma^2_{\epsilon, 1, i}(1) + p_1^\mathcal{J} \sigma^2_{\epsilon, 1, i}(0) ) \Pr(A_{1(i)} = 1) }{\left(\frac{1}{n}\sumi \bbE[M_i^1 - M_i^0] \right)^2 p_1^\mathcal{J} p_0^\mathcal{J}}\right),
\end{align}
where $\sigma^2_{\epsilon, 1, i}(d) \coloneqq \bbE[\epsilon_{1,i}^2 \mid D_{1(i)} = d, A_{1(i)} = 1]$.
We can see that the asymptotic variances of the 2SLS and WLS estimators have the same denominator term, while the numerator of the WLS is weighted by the compliance probability $\Pr(A_{1(i)} = 1)$.
Thus, if $\sigma^2_{\varepsilon,i}(d)$ and $\sigma^2_{\varepsilon,1,i}(d)$ take close values, the spillover-effect parameter can be estimated more efficiently using the WLS estimator.

For derivation of \eqref{eq:clt_net_2sls} and \eqref{eq:clt_net_sls}, see Appendix \ref{app:proof_sec3}.

\section{Statistical inference}\label{sec:inference}

The methods presented in the preceding sections may be difficult to apply directly in practice because of some restrictive assumptions regarding the data structure.
In addition, we need to deal with the heterogeneity in the distribution of population residuals for statistical inference. 
Considering these points, we present several empirically tractable inference procedures by introducing additional constraints that may or may not be plausible depending on the data at hand.

\subsection{Subset analysis}

To utilise the asymptotic normality results shown above, we require the population residuals to have zero mean uniformly.
One possible way to enforce this property is to limit our attention to a subset in which both the potential outcomes and network structures are IID.
The units in this subset must have potential peers of the same size.
A similar approach is considered in \cite{vazquez2022identification}.
If we can find such a subset, say $\mathcal{I}'$, then the population 2SLS estimand can be defined as
\begin{align*}
    (\beta_0^{2sls}, \beta_d^{2sls}, \beta_s^{2sls}) = \argmin_{b_0, b_d, b_s} \bbE[(Y_i - b_0 - b_d D_i - b_s \bbL(M_i \mid D_i, D_{1(i)}))^2 \mid i \in \mathcal{I}'],
\end{align*}
 where $\bbL(M_i \mid D_i, D_{1(i)}) \coloneqq \gamma_0 + \gamma_d D_i + \gamma_s D_{1(i)}$, and $(\gamma_0, \gamma_d, \gamma_s) = \argmin_{a_0, a_d, a_s} \bbE[(M_i - a_0 - a_d D_i - a_s D_{1(i)})^2 \mid i \in \mathcal{I}']$.
For simplicity, the dependence of the parameters on the choice of $\mathcal{I}'$ is suppressed.
Then, for each $i \in \mathcal{I}'$, we can show that $\bbE[\varepsilon_i \mid i \in \mathcal{I}'] = \bbE[D_i \varepsilon_i \mid i \in \mathcal{I}'] = \bbE[D_{1(i)} \varepsilon_i \mid i \in \mathcal{I}'] = 0$ holds for the population residual $\varepsilon_i = Y_i - \beta_0^{2sls} - D_i \beta_d^{2sls} - M_i \beta_s^{2sls}$.
Moreover, the asymptotic distribution of the 2SLS estimator is simplified as
\begin{align*}
    \sqrt{n'}(\hat \beta_s^{2sls} - \beta_s^{2sls}) \overset{d}{\to} N\left(0, \frac{ p_0^\mathcal{J} \sigma^2_\varepsilon (1) + p_1^\mathcal{J} \sigma^2_\varepsilon (0)}{ \bbE[M_i^1 - M_i^0 \mid A_{1(i)} = 1, i \in \mathcal{I}']^2 (q_1^A)^2 p_1^\mathcal{J} p_0^\mathcal{J}} \right),
\end{align*}
where $n' \coloneqq |\mathcal{I}'|$, $\sigma^2_\varepsilon (d) \coloneqq \bbE [\varepsilon_i^2 \mid D_{1(i)} = d, i \in \mathcal{I}']$, and $q_d^A \coloneqq \Pr(A_{1(i)} = 1 \mid i \in \mathcal{I}')$.
A similar result is obtained for the WLS estimator:
\begin{align*}
    \sqrt{n'}(\hat \beta_s^{wls} - \beta_s^{wls}) \overset{d}{\to} N\left(0, \frac{ p_0^\mathcal{J} \sigma^2_{\epsilon,1} (1) + p_1^\mathcal{J} \sigma^2_{\epsilon,1} (0)}{ \bbE[M_i^1 - M_i^0 \mid A_{1(i)} = 1, i \in \mathcal{I}']^2 q_1^A p_1^\mathcal{J} p_0^\mathcal{J}} \right),
\end{align*}
where $\sigma^2_{\epsilon,1} (d) \coloneqq \bbE [\epsilon_{1,i}^2 \mid D_{1(i)} = d, A_{1(i)} = 1, i \in \mathcal{I}']$.
See Appendix \ref{app:proof_sec4} for the derivation of these results.

\subsection{Homogeneous treatment effects model}

Another possible approach, as in many empirical studies, is to assume a homogeneous treatment effect model:
\begin{align}\label{eq:homo_model}
    Y_i = \beta_0 + \beta_d D_i + \beta_s M_i + \varepsilon_i, \;\; \bbE[\varepsilon_i] = 0, \;\; \text{for} \;\; i \in \mathcal{I}.     
\end{align}
Here, we implicitly require that the specification of the exposure mapping $M_i(\cdot)$ be correct (otherwise, the misspecification error may correlate with IV).
There are many advantages of this approach: we do not need to account for the heterogeneity in the data structure, the treatment assignment does not have to follow a simple Bernoulli design, we can incorporate $\mathcal{J}$ in the estimation as well, the IV does not have to be binary, etc.
One clear limitation is that it is often difficult to assume treatment homogeneity in practical applications.

\subsection{Randomization inference}\label{subsec:frt}

If one simply wants to test for the presence of spillover effects within the data at hand, the Fisher's randomization test is a promising alternative.
For example, consider the following null hypothesis:
\begin{align*}
    \mathbb{H}_0: \tau_i^1(d,r) = 0 \;\; \text{for all} \;\; (d,r) \in \{0,1\} \times \{1, \ldots, n_i\}, \; i \in \mathcal{I}.
\end{align*}
If $\mathbb{H}_0$ is true, the IV has no impact on the outcome ($\beta_s^{2sls} = \beta_s^{wls} = 0$) so that we can impute the values of all $\{Y_i(D_i, R_i(d))\}_{d \in \{0,1\}}$ as $Y_i(D_i, R_i)$ ($=Y_i$).
Thus, we can consider a conditional randomization test in which $\bm{D}_{1,n} = (D_{1(1)}, \ldots, D_{1(n)})^\top$ are randomized and everything else is fixed.
Specifically, letting $T(\bm{D}_{1,n})$ be a predetermined test statistic, we can approximate the $p$-value of the statistic in the following manner: (\textit{Step 1}) Compute $T(\bm{D}_{1,n})$; (\textit{Step 2}) Draw $\bm{d}_{1,n}^{(b)}$ independently from the appropriate conditional distribution of $\bm{D}_{1,n}$ and compute $T(\bm{d}_{1,n}^{(b)})$ for $b = 1, \ldots , B$; and (\textit{Step 3}) Compute $\hat p_B \coloneqq B^{-1} \sum_{b=1}^B \bm{1}\{T(\bm{d}_{1,n}^{(b)}) \ge T(\bm{D}_{1,n})\}$.

Two obvious candidates for the test statistic $T(\bm{D}_{1,n})$ are $\hat \beta_s^{2sls}$ and $\hat \beta_s^{wls}$ possibly with some normalization.
For other options, we can consider the intention-to-treat (ITT) statistic and ITT for compliers (ITTC):
\begin{align*}
    T^\text{ITT}(\bm{D}_{1,n}) 
    & = \frac{\sumi D_{1(i)} Y_i}{\sumi D_{1(i)}} - \frac{\sumi (1 - D_{1(i)}) Y_i}{\sumi (1 - D_{1(i)})} \\
    T^\text{ITTC}(\bm{D}_{1,n})
    & = \frac{\sumi A_{1(i)} D_{1(i)} Y_i}{\sumi A_{1(i)} D_{1(i)}} - \frac{\sumi A_{1(i)} (1 - D_{1(i)}) Y_i}{\sumi A_{1(i)} (1 - D_{1(i)})}.
\end{align*}
Similar randomization tests can be found in the literature (e.g., \citealp{rogowski2012estimating, forastiere2018posterior, kang2018inference}).
Note that to perform the randomization test, the experimental design does not have to be a Bernoulli design as long as it is known (although the above causal interpretations may not hold).

\section{Empirical Illustration}\label{sec:empiric}

In this section, we revisit the data from \citet{paluck2016changing}, who conducted a large social experiment on anti-conflict intervention programs in American middle schools.
Half of these schools were randomly selected to host the programs.
Within each selected school, a group of students ({\it seed-eligible students}) were selected and half of them ({\it seed students}) were randomly invited to join the meeting program.
The students' social networks were measured by asking them to nominate up to 10 closest friends in their school.

In our analysis, the treatment variable $D_i$ indicates whether student $i$ was a seed student.\footnote{
    Participation in the anti-conflict intervention program was not mandatory for seed students.
    Therefore, what we are estimating here is the effect of being nominated as a seed student.
}
We assume identity mapping for $M_i(\cdot)$; that is, $M_i = R_i$.
For the IV of $M_i$, we use $D_{1(i)}$, where $1(i)$ denotes the closest seed-eligible friend to $i$.
Outcome $Y$ is a dummy indicator for wearing the program wristband given as a reward for engaging in conflict-mitigating behaviors.

Under this setup, we consider the following two samples: (1) a sample of students in treatment schools who have at least one seed-eligible friend, and (2) a subsample of (1) constructed so that the friend networks of the units are non-overlapping.
We perform OLS and 2SLS estimations on these samples. 
Because every unit has a seed-eligible friend (i.e., $A_{1(i)} = 1$ for all $i$), the 2SLS and WLS estimators are equivalent.
In sample (1), we do not rule out the possibility that the units in this sample share the same friends.
While this is not consistent with the theoretical analysis part, we examine this sample for robustness checks.
Sample (2) is the one that conforms with the theoretical framework considered above.
Note that the formation of potential peers (i.e., the set of general schoolmates) is treated as a given factor, and thus any school-level endogeneity may not be isolated from the estimates (see also the discussion in footnote 3).

The results are summarized in Table \ref{table:result}.
The spillover effect is significant in all four cases, irrespective of the estimation method; having one additional treated friend increases the probability of receiving the reward by roughly 5\% or so.
As a robustness check, we also perform the randomization test proposed in Subsection \ref{subsec:frt} using sample (2).
The $p$-values obtained from the 2SLS and ITT statistics are 0.010 and 0.016, respectively.
From these results, it would be safe to conclude that there exist spillover effects of the program.

\begin{table}[ht]
    \caption{Estimation results}
    \label{table:result}
    \centering
    \begin{tabular}{l|cccccccc}
        \hline\hline
         & \multicolumn{2}{c}{OLS} & \multicolumn{2}{c}{2SLS} & \multicolumn{2}{c}{OLS} & \multicolumn{2}{c}{2SLS} \\
        Variable & Coef. & t-value & Coef. & t-value & Coef. & t-value & Coef. & t-value \\
        \hline
        $D$ & 0.122 & 6.620 & 0.122 & 6.577 & 0.255 & 4.034 & 0.257 & 4.062 \\
        $M$ & 0.041 & 6.333 & 0.034 & 3.547 & 0.054 & 2.452 & 0.073 & 2.284 \\
        Sample & \multicolumn{2}{c}{(1)} & \multicolumn{2}{c}{(1)} & \multicolumn{2}{c}{(2)} & \multicolumn{2}{c}{(2)} \\
        \hline
    \end{tabular}

    \footnotesize
    (1) Students in treatment schools with at least one seed-eligible friend (size = 6603)

	(2) Subsample of (1) such that $\mathcal{P}_i$ and $\mathcal{P}_{i'}$ are disjoint for any $i \neq i'$ (size = 562) 
\end{table}

\clearpage
\bibliography{references.bib}

\clearpage

\appendix
\def \baselinestretch{1.2} 
\setcounter{page}{1}
\small
\begin{center}
\LARGE Appendix (not for publication)
\end{center}

\section{Proofs}\label{app:proof}

In the following, we introduce the re-parameterization $\beta_1 \coloneqq \beta_d + \beta_0$.
In addition, when there is no confusion, superscripts (such as \textit{ols}, \textit{2sls}, and \textit{wls}) are suppressed for simplicity.

\subsection{Proofs of results in Section \ref{sec:pair}}\label{app:proof_sec2}

\begin{flushleft}
\textbf{Proof of Proposition \ref{prop:pair_ls}}
\end{flushleft}

(i) The objective function is written as 
\begin{align*}
    \bbE[(Y_i - b_0(1- D_i) - b_1 D_i - b_s S_i)^2]  = \bbE[(Y_i - b_1 - b_s S_i)^2 D_i ] + \bbE[(Y_i - b_0 - b_s S_i)^2 (1 - D_i) ].
\end{align*}
Using simple algebra, we obtain $\beta_0 = \bbE[Y_i \mid D_i = 0] - \beta_s \bbE[S_i]$ and $\beta_1  = \bbE[Y_i \mid D_i = 1] - \beta_s \bbE[S_i]$, which leads to $\beta_d = \bbE[Y_i \mid D_i = 1] - \bbE[Y_i \mid D_i = 0]$.
Then, the result follows from 
\begin{align*}
    \bbE[Y_i \mid D_i = d]
    & = \bbE[ Y_i(d, 1) \mid A_{j(i)}= 1] p_1^A p_1^\mathcal{J} + \bbE[ Y_i(d, 0) \mid A_{j(i)}= 1] 
 p_1^A p_0^\mathcal{J} + \bbE[ Y_i(d, 0) \mid A_{j(i)}= 0 ] p_0^A.
\end{align*}

(ii) The first-order condition with respect to $\beta_s$ is
\begin{align*}
    0  
    & = \bbE[S_i(Y_i - \beta_0(1 - D_i) - \beta_1 D_i - \beta_s S_i) ] \\
    & = \bbE[S_i(Y_i - \beta_1 - \beta_s S_i) D_i ] +  \bbE[S_i(Y_i - \beta_0 - \beta_s S_i) (1 - D_i)].
\end{align*}
Observe that
\begin{align*}
    \bbE[S_i(Y_i - \beta_1 - \beta_s S_i) D_i ] 
    & = \bbE[S_i (Y_i - \bbE[Y_i \mid D_i = 1] + \beta_s \bbE[S_i] - \beta_s S_i) D_i ] \\
    & = \bbE[S_i(Y_i - \bbE[Y_i \mid D_i = 1])D_i] - \beta_s \bbE[ (S_i - S_i \bbE[S_i]) D_i] \\
    & =  \Cov(S_i, Y_i \mid D_i = 1) p_1^\mathcal{I} - \beta_s \Var(S_i) p_1^\mathcal{I}.
\end{align*}
Similarly, $\bbE[S_i(Y_i - \beta_0 - \beta_s S_i) (1 - D_i) ] = \Cov(S_i, Y_i \mid D_i = 0) p_0^\mathcal{I} - \beta_s \Var(S_i) p_0^\mathcal{I}$.
Combining these yields
\begin{align*}
    \beta_s = \frac{\Cov(S_i, Y_i \mid D_i = 1) p_1^\mathcal{I} + \Cov(S_i, Y_i \mid D_i = 0) p_0^\mathcal{I} }{\Var(S_i)}.
\end{align*}
Here, observe that
\begin{align*}
    \Cov(S_i, Y_i \mid D_i = d) 
    & = \bbE[S_i Y_i \mid D_i = d] - \bbE[S_i \mid D_i = d]\bbE[Y_i \mid D_i = d] \\
    & = \bbE[S_i] \bbE[Y_i \mid S_i = 1, D_i = d] - \bbE[S_i]^2 \bbE[Y_i \mid S_i = 1, D_i = d] \\
    & \quad - \bbE[S_i](1 - \bbE[S_i]) \bbE[Y_i \mid S_i = 0, D_i = d] \\
    & = \{\bbE[Y_i \mid S_i = 1, D_i = d] - \bbE[Y_i \mid S_i = 0, D_i = d]\} \Var(S_i).
\end{align*}
Moreover,
\begin{align*}
    & \bbE[Y_i \mid S_i = 1, D_i = d] - \bbE[Y_i \mid S_i = 0, D_i = d] \\
    & = \bbE[Y_i \mid A_{j(i)}= 1, D_{j(i)} = 1, D_i = d] - \bbE[Y_i \mid A_{j(i)}= 1, D_{j(i)} = 0, D_i = d]\frac{p_1^A p_0^\mathcal{J}}{1 - p_1^A p_1^\mathcal{J}} \\
    & \quad - \bbE[Y_i \mid A_{j(i)}= 0, D_i = d]\frac{p_0^A}{1 - p_1^A p_1^\mathcal{J}} \\
    & = \bbE[Y_i(d, 1) \mid A_{j(i)}= 1] - \bbE[Y_i(d,0) \mid A_{j(i)}= 1] \frac{p_1^A p_0^\mathcal{J}}{1 - p_1^A p_1^\mathcal{J}} - \bbE[Y_i(d, 0) \mid A_{j(i)}= 0] \frac{p_0^A}{1 - p_1^A p_1^\mathcal{J}} \\
    & = \bbE[\tau_i(d) \mid A_{j(i)}= 1] + \eta_i(d) \frac{p_0^A}{1 - p_1^A p_1^\mathcal{J}}.
\end{align*}
Thus,
\begin{align*}
    \beta_s
    & = \sum_{d \in \{0,1\}} \left\{ \bbE[Y_i \mid S_i = 1, D_i = d] - \bbE[Y_i \mid S_i = 0, D_i = d] \right\} p_d^\mathcal{I} \\
    & = \sum_{d \in \{0,1\}} \left\{ \bbE[\tau_i(d) \mid A_{j(i)}= 1] + \eta_i(d) \frac{p_0^A}{1 - p_1^A p_1^\mathcal{J}}\right\}p_d^\mathcal{I},
\end{align*}
as desired.
\qed

\bigskip

\begin{flushleft}
\textbf{Proof of Proposition \ref{prop:pair_2sls}}
\end{flushleft}

From \eqref{eq:pair_2sls_exp}, we can see that
\begin{align*}
    \beta_s = \frac{\Cov(D_{j(i)}, Y_i \mid D_i = 1)p_1^\mathcal{I} + \Cov(D_{j(i)}, Y_i \mid D_i = 0)p_0^\mathcal{I}}{\Cov(D_{j(i)}, S_i)}.
\end{align*}
We observe that $\Cov(D_{j(i)}, Y_i \mid D_i = d) = \{\bbE[Y_i \mid D_{j(i)} = 1, D_i = d] - \bbE[Y_i \mid D_{j(i)} = 0, D_i = d]\}  p_1^\mathcal{J} p_0^\mathcal{J}$, 
\begin{align*}
    \bbE[Y_i \mid D_{j(i)} = 1, D_i = d] 
    & = \bbE[Y_i \mid A_{j(i)}= 1, D_{j(i)} = 1, D_i = d]p_1^A + \bbE[Y_i \mid A_{j(i)}= 0, D_{j(i)} = 1, D_i = d]p_0^A \\
    & = \bbE[Y_i(d,1) \mid A_{j(i)}= 1] p_1^A + \bbE[Y_i(d,0) \mid A_{j(i)}= 0]p_0^A,
\end{align*}
and similarly,
\begin{align*}
    \bbE[Y_i \mid D_{j(i)} = 0, D_i = d] 
    & = \bbE[Y_i(d,0) \mid A_{j(i)}= 1]p_1^A + \bbE[Y_i(d,0) \mid A_{j(i)}= 0]p_0^A.
\end{align*}
Therefore,
\begin{align*}
    \Cov(D_{j(i)}, Y_i \mid D_i = d )
    & = \bbE[\tau_i(d) \mid A_{j(i)}= 1]p_1^A p_1^\mathcal{J} p_0^\mathcal{J}.
\end{align*}
Noting that $\Cov(D_{j(i)}, S_i) = \bbE[A_{j(i)} D_{j(i)}] - \bbE[A_{j(i)} D_{j(i)}]\bbE[D_{j(i)}] = p_1^A p_1^\mathcal{J} p_0^\mathcal{J}$, we obtain the desired result.
\qed


\bigskip

\begin{flushleft}
\textbf{Proof of Proposition \ref{prop:pair_sel_ls}}
\end{flushleft}

(i) The objective function can be rewritten as 
\begin{align*}
    \bbE[A_{j(i)} (Y_i - b_0(1- D_i) - b_1 D_i - b_s D_{j(i)})^2] 
    & = \bbE[A_{j(i)} (Y_i - b_1 - b_s D_{j(i)})^2 D_i ] \\
    & \quad + \bbE[A_{j(i)} (Y_i - b_0 - b_s D_{j(i)})^2 (1 - D_i) ].
\end{align*}
By solving the first-order condition for $\beta_1$, we obtain
\begin{align*}
    0 & = \bbE[Y_i \mid D_i = 1, A_{j(i)}= 1] - \beta_1 - \beta_s \bbE[ D_{j(i)} \mid D_i = 1, A_{j(i)}= 1],
\end{align*}
which leads to $\beta_1  = \bbE[Y_i \mid D_i = 1, A_{j(i)}= 1] - \beta_s p_1^\mathcal{J}$.
Similarly, $\beta_0 = \bbE[Y_i \mid D_i = 0, A_{j(i)}= 1] - \beta_s p_1^\mathcal{J}$.
Hence, $\beta_d = \bbE[Y_i \mid D_i = 1, A_{j(i)}= 1] - \bbE[Y_i \mid D_i = 0, A_{j(i)}= 1]$.
Since   
\begin{align*}
    \bbE[Y_i \mid D_i = d, A_{j(i)}= 1]
    & = \bbE[ Y_i \mid D_i = d, A_{j(i)}= 1, D_{j(i)} = 1] p_1^\mathcal{J} + \bbE[ Y_i \mid D_i = d, A_{j(i)}= 1, D_{j(i)} = 0] p_0^\mathcal{J} \\
    & = \bbE[ Y_i(d, 1) \mid A_{j(i)}= 1] p_1^\mathcal{J} + \bbE[ Y_i(d, 0) \mid A_{j(i)}= 1] p_0^\mathcal{J},
\end{align*}
(i) is proved.

\bigskip

(ii) By combining the first-order condition for $\beta_s$ and result (i), we can easily show that
\begin{align*}
    \beta_s = \frac{\Cov(D_{j(i)}, Y_i \mid  D_i = 1, A_{j(i)}= 1) p_1^\mathcal{I} + \Cov(D_{j(i)}, Y_i \mid  D_i = 0, A_{j(i)}= 1) p_0^\mathcal{I} }{\Var(D_{j(i)})}.
\end{align*}
Moreover, observe that
\begin{align*}
    & \Cov(D_{j(i)}, Y_i \mid  D_i = d, A_{j(i)}= 1) \\
    & = \bbE[D_{j(i)} Y_i \mid D_i = d, A_{j(i)}= 1] - \bbE[D_{j(i)} \mid D_i = d, A_{j(i)}= 1]\bbE[Y_i \mid D_i = d, A_{j(i)}= 1] \\
    & = p_1^\mathcal{J} \bbE[Y_i \mid S_i = 1, D_i = d] - (p_1^\mathcal{J})^2 \bbE[Y_i \mid S_i = 1, D_i = d] - p_1^\mathcal{J} p_0^\mathcal{J} \bbE[Y_i \mid D_{j(i)} = 0, D_i = d, A_{j(i)}= 1] \\
    & = \{\bbE[Y_i \mid S_i = 1, D_i = d] - \bbE[Y_i \mid D_{j(i)} = 0, D_i = d, A_{j(i)}= 1]\} \Var(D_{j(i)}) \\
    & = \{\bbE[Y_i(d,1) \mid A_{j(i)}= 1] - \bbE[Y_i(d,0) \mid A_{j(i)}= 1]\} \Var(D_{j(i)}).
\end{align*}
This completes the proof.
\qed


\bigskip

\begin{flushleft}
\textbf{Derivation of the asymptotic normality results}
\end{flushleft}

Let $\bm{D}_{J,n} = (D_{j(1)}, \ldots, D_{j(n)})^\top$, $\bm{S}_n = (S_1, \ldots, S_n)^\top$,  $\tilde{\bm{D}}_{J,n} \coloneqq \bm{D}_{J,n} - \bm{D}_{c,n}(\bm{D}_{c,n}^\top \bm{D}_{c,n})^{-1}\bm{D}_{c,n}^\top \bm{D}_{J,n}$, $\mathbb{I}_{n, A} \coloneqq \text{diag}(A_{j(1)}, \ldots, A_{j(n)})$, and $\tilde{\bm{D}}_{J,n,A} \coloneqq \mathbb{I}_{n, A}\bm{D}_{J,n} - \mathbb{I}_{n, A}\bm{D}_{c,n}(\bm{D}_{c,n}^\top \mathbb{I}_{n, A} \bm{D}_{c,n})^{-1}\bm{D}_{c,n}^\top \mathbb{I}_{n, A}\bm{D}_{J,n}$.
Then, the sample version of \eqref{eq:pair_2sls_exp} is obtained by $\hat \beta_s^{2sls} = \tilde{\bm{D}}_{J,n}^\top \bm{Y}_n/ \tilde{\bm{D}}_{J,n}^\top \bm{S}_n$.
Similarly, the sample analog of $\beta_s^{wls}$ is given by $\hat \beta_s^{wls} = \tilde{\bm{D}}_{J,n,A}^\top \bm{Y}_n/ \tilde{\bm{D}}_{J,n,A}^\top \bm{D}_{J,n}$.

In the following, we discuss only the normality result for the WLS estimator.
Observe that
\begin{align*}
    \sqrt{n}(\hat \beta_s^{wls} - \beta_s^{wls})
    & = \sqrt{n}( \tilde{\bm{D}}_{J,n,A}^\top (\bm{D}_{c,n} \beta_c^{wls} + \bm{D}_{J,n} \beta_s^{wls} + \bm{\epsilon}_{1,n} ) / \tilde{\bm{D}}_{J,n,A}^\top \bm{D}_{J,n} - \beta_s^{wls})\\
    & = \sqrt{n} \tilde{\bm{D}}_{J,n,A}^\top \bm{\epsilon}_{1,n} / \tilde{\bm{D}}_{J,n,A}^\top \bm{D}_{J,n},
\end{align*}
where $\bm{\epsilon}_{1,n} = (\epsilon_{1,1}, \ldots, \epsilon_{1,n})^\top$ and $\beta_c^{wls} = (\beta_0^{wls}, \beta_d^{wls})^\top$.
Note that the population residual has the following form: $\epsilon_i = A_{j(i)}\epsilon_{1, i}$, where
    \begin{align*}
        \epsilon_{1,i}
        & = Y_i - D_i \bbE[Y_i \mid D_i = 1, A_{j(i)} = 1] - (1 - D_i) \bbE[Y_i \mid D_i = 0, A_{j(i)} = 1] -  \beta_s^{wls} ( D_{j(i)} - p_1^\mathcal{J}) \\
        & = D_i \{Y_i(1,1)D_{j(i)} - \bbE[Y_i(1,1)D_{j(i)} \mid A_{j(i)} = 1]\} \\
        & \quad + D_i \{ Y_i(1,0)(1 - D_{j(i)}) - \bbE[Y_i(1,0)(1 - D_{j(i)}) \mid A_{j(i)}= 1] \} \\
        & \quad + (1 - D_i) \{ Y_i(0,1)D_{j(i)} - \bbE[Y_i(0,1)D_{j(i)} \mid A_{j(i)}= 1]\} \\
        & \quad + (1 - D_i) \{ Y_i(0,0)(1 - D_{j(i)}) - \bbE[Y_i(0,0)(1 - D_{j(i)}) \mid A_{j(i)}= 1] \} - \beta_s^{wls} ( D_{j(i)} - p_1^\mathcal{J})
\end{align*}
for all $i$ such that $A_{j(i)} = 1$.
By direct calculations, we can confirm that $\bbE[\epsilon_i] = \bbE[D_i \epsilon_i] = \bbE[D_{j(i)} \epsilon_i] = 0$ holds.\footnote{
   For the 2SLS estimand, the population residual is obtained as follows:
    \begin{align*}
        \varepsilon_i 
        & = Y_i - D_i \bbE[Y_i \mid D_i = 1] - (1 - D_i) \bbE[Y_i \mid D_i = 0] - \beta_s^{2sls} (S_i - \bbE[S_i]) \\
        & = D_i\{Y_i(1,1)S_i - \bbE[Y_i(1,1)S_i] + Y_i(1,0)(1 - S_i) - \bbE[Y_i(1,0)(1 - S_i)]\} \\
        & \quad + (1 - D_i)\{Y_i(0,1)S_i - \bbE[Y_i(0,1)S_i] + Y_i(0,0)(1 - S_i) - \bbE[Y_i(0,0)(1 - S_i)]\} - \beta_s^{2sls}(S_i - p_1^\mathcal{J} p_1^A).
    \end{align*}
     Direct calculations yield that $\bbE[\varepsilon_i] = \bbE[D_i \varepsilon_i] = \bbE[D_{j(i)} \varepsilon_i] = 0$.
}
Markov's inequality with Assumptions \ref{as:rct_pair} and \ref{as:iid_pair} yields that  $\bm{D}_{c,n}^\top \mathbb{I}_{n, A} \bm{D}_{c,n}/n = p_1^A \left(\begin{array}{cc} 1 & p_1^\mathcal{I} \\ p_1^\mathcal{I} & p_1^\mathcal{I} \end{array}\right) + O_P(n^{-1/2})$ and $\bm{D}_{c,n}^\top \mathbb{I}_{n, A} \bm{D}_{J,n}/n = p_1^A \left(\begin{array}{c} p_1^\mathcal{J} \\ p_1^\mathcal{I} p_1^\mathcal{J} \end{array}\right) + O_P(n^{-1/2})$, leading to $(\bm{D}_{c,n}^\top \mathbb{I}_{n, A} \bm{D}_{c,n}/n)^{-1} \bm{D}_{c,n}^\top \mathbb{I}_{n, A} \bm{D}_{J,n}/n = (p_1^\mathcal{J}, 0)^\top + o_P(1)$.
Then, for the denominator on the right-hand side, direct calculations yield $\tilde{\bm{D}}_{J,n,A}^\top \bm{D}_{J,n}/n = p_1^A p_1^\mathcal{J} p_0^\mathcal{J} + o_P(1)$.
For the numerator,
\begin{align*}
    \tilde{\bm{D}}_{J,n,A}^\top \bm{\epsilon}_{1,n}/ \sqrt{n} 
    & = \frac{1}{\sqrt{n}} \sumi D_{j(i)} \epsilon_i - (\bm{D}_{J,n}^\top \mathbb{I}_{n, A} \bm{D}_{c,n}/n)(\bm{D}_{c,n}^\top \mathbb{I}_{n, A} \bm{D}_{c,n}/n)^{-1} \left( \frac{1}{\sqrt{n}} \sumi D_{c,i}\epsilon_i\right)\\
    & = \frac{1}{\sqrt{n}} \sumi (D_{j(i)} - p_1^\mathcal{J}) A_{j(i)} \epsilon_{1, i} + o_P(1),
\end{align*}
where the last equality follows from $||n^{-1/2} \sumi D_{c,i}\epsilon_i|| = O_P(1)$ under the independence and a bounded second moment assumption.
Further, 
\begin{align*}
    & \Var\left(\frac{1}{\sqrt{n}} \sumi (D_{j(i)} - p_1^\mathcal{J}) A_{j(i)} \epsilon_{1, i}\right) \\
    & = \bbE\left[(D_{j(i)} - p_1^\mathcal{J})^2 A_{j(i)} \epsilon_{1, i}^2 \right] \\
    & = (p_0^\mathcal{J})^2 p_1^\mathcal{J} p_1^A \bbE\left[\epsilon_{1, i}^2 \mid A_{j(i)}= 1, D_{j(i)} = 1 \right] + p_0^\mathcal{J} (p_1^\mathcal{J})^2 p_1^A \bbE\left[\epsilon_{1, i}^2 \mid A_{j(i)}= 1, D_{j(i)} = 0 \right].
\end{align*}
Finally, the result follows from the central limit theorem and Slutsky's theorem:
\begin{align*}
    \sqrt{n}(\hat \beta_s^{wls} - \beta_s^{wls})
    \overset{d}{\to} N \left(0, \frac{p_0^\mathcal{J}\bbE[\epsilon_{1, i}^2 \mid A_{j(i)}= 1, D_{j(i)} = 1] + p_1^\mathcal{J}\bbE[\epsilon_{1, i}^2 \mid A_{j(i)}= 1, D_{j(i)} = 0]}{p_1^A p_1^\mathcal{J} p_0^\mathcal{J}}\right).
\end{align*}
\qed


\subsection{Proofs of results in Section \ref{sec:gen}}\label{app:proof_sec3}

\begin{flushleft}
\textbf{Proof of Theorem \ref{thm:net_ls}}
\end{flushleft}

(i) The objective function can be rewritten as 
\begin{align*}
    \sumi\bbE[(Y_i - b_0(1- D_i) - b_1 D_i - b_s M_i)^2]  = \sumi\bbE[(Y_i - b_1 - b_s M_i)^2 D_i ] + \sumi\bbE[(Y_i - b_0 - b_s M_i)^2 (1 - D_i) ].
\end{align*}
By solving the first-order condition for $\beta_1$, we obtain
\begin{align*}
    0 = \sumi\bbE[Y_i \mid D_i = 1] - n \beta_1 - \beta_s \sumi\bbE[ M_i \mid D_i = 1 ],
\end{align*}
which leads to $\beta_1 = n^{-1} \sumi\bbE[Y_i \mid D_i = 1] - \beta_s \mu$.
Similarly, $\beta_0 = n^{-1} \sumi\bbE[Y_i \mid D_i = 0] - \beta_s \mu$.
Thus, $\beta_d = n^{-1}\sumi(\bbE[Y_i \mid D_i = 1] - \bbE[Y_i \mid D_i = 0])$.
Then, the result follows from
\begin{align*}
    \bbE[Y_i \mid D_i = d]
    & = \sum_{(r, \vec{a}) \in \mathcal{R}_i \times \mathcal{A}_i} \bbE[Y_i \mid R_i = r, \bm{A}_{\mathcal{P}_i} = \vec{a}, D_i = d] \Pr(R_i = r, \bm{A}_{\mathcal{P}_i} = \vec{a}) \\
    & = \sum_{(r, \vec{a}) \in \mathcal{R}_i \times \mathcal{A}_i} \bbE[Y_i \mid \vec{a}^\top \bm{D}_{\mathcal{P}_i} = r, \bm{A}_{\mathcal{P}_i} = \vec{a}, D_i = d] \Pr(R_i = r, \bm{A}_{\mathcal{P}_i} = \vec{a}) \\
    & = \sum_{(r, \vec{a}) \in \mathcal{R}_i \times \mathcal{A}_i} \bbE[Y_i(d,r) \mid \bm{A}_{\mathcal{P}_i} = \vec{a}] \Pr(R_i = r, \bm{A}_{\mathcal{P}_i} = \vec{a}).
\end{align*}

\bigskip

(ii) The first-order condition with respect to $\beta_s$ yields
\begin{align*}
    0  
    & = \sumi\bbE[M_i(Y_i - \beta_0 - \beta_d D_i - \beta_s M_i) ] \\
    & = \sumi\bbE\left[M_i\left(Y_i - \frac{1}{n} \sumj \bbE[Y_j \mid D_j = 0](1 - D_i) - \frac{1}{n} \sumj \bbE[Y_j \mid D_j = 1] D_i  - \beta_s [M_i - \mu] \right) \right] \\
    & = \sumi\bbE\left[M_i Y_i \right] - \mu \sumi \left(  \bbE[Y_i \mid D_i = 0] p_0^\mathcal{I} + \bbE[Y_i \mid D_i = 1] p_1^\mathcal{I} \right) - \beta_s \sumi (\bbE[M_i^2] - \bbE[M_i] \mu) \\
    & = \sumi\bbE\left[ (M_i - \mu) Y_i \right] - \beta_s \sumi \bbE\left[ (M_i - \mu)^2 \right].
\end{align*}
Therefore,
\begin{align*}
    \beta_s = \frac{ \sumi \left( \bbE \left[(M_i - \mu) Y_i \mid D_i = 1 \right] p_1^\mathcal{I} + \bbE \left[(M_i - \mu)  Y_i \mid D_i = 0 \right] p_0^\mathcal{I}  \right) }{\sumi \bbE\left[ (M_i - \mu)^2 \right] }.
\end{align*}
Here, observe that
\begin{align*}
   \bbE \left[(M_i - \mu) Y_i \mid D_i = d \right]
   & = \sum_{r \in \mathcal{R}_i} (M_i(r) - \mu) \Pr(R_i = r) \bbE[Y_i \mid R_i = r, D_i = d],
\end{align*}
and moreover
\begin{align*}
    \bbE[Y_i \mid R_i = r, D_i = d] 
    & = \sum_{\vec{a} \in \mathcal{A}_i} \bbE[Y_i(d, r) \mid \vec{a}^\top\bm{D}_{\mathcal{P}_i} = r, \bm{A}_{\mathcal{P}_i} = \vec{a}, D_i = d] \Pr(\bm{A}_{\mathcal{P}_i} = \vec{a} \mid R_i = r) \\
    & = \sum_{\vec{a} \in \mathcal{A}_i} \bbE[Y_i(d, r) \mid \bm{A}_{\mathcal{P}_i} = \vec{a}] \Pr(\bm{A}_{\mathcal{P}_i} = \vec{a} \mid R_i = r) \\
    & = \sum_{\vec{a} \in \mathcal{A}_i} \bbE[\tau_i^0(d, r) \mid \bm{A}_{\mathcal{P}_i} = \vec{a}] \Pr(\bm{A}_{\mathcal{P}_i} = \vec{a} \mid R_i = r) \\
    & \quad + \sum_{\vec{a} \in \mathcal{A}_i}\bbE[Y_i(d, 0) \mid \bm{A}_{\mathcal{P}_i} = \vec{a}] \Pr(\bm{A}_{\mathcal{P}_i} = \vec{a} \mid R_i = r).
\end{align*}
Thus,
\begin{align*}
    \beta_s
    & = \frac{\sumi \sum_{(r, \vec{a}) \in \mathcal{R}_i \times \mathcal{A}_i} (M_i(r) - \mu)\Pr(R_i = r, \bm{A}_{\mathcal{P}_i} = \vec{a}) \sum_{d \in \{0,1\}} \bbE[\tau_i^0(d, r) \mid \bm{A}_{\mathcal{P}_i} = \vec{a}] p_d^\mathcal{I}}{\sumi \bbE\left[ (M_i - \mu)^2 \right]} \\
    & \quad + \frac{\sumi \sum_{(r, \vec{a}) \in \mathcal{R}_i \times \mathcal{A}_i}  (M_i(r) - \mu) \Pr(R_i = r, \bm{A}_{\mathcal{P}_i} = \vec{a}) \sum_{d \in \{0,1\}}  \bbE[Y_i(d,0) \mid \bm{A}_{\mathcal{P}_i} = \vec{a}] p_d^\mathcal{I}}{\sumi \bbE\left[ (M_i - \mu)^2 \right]}.
\end{align*}
This completes the proof.
\qed


\bigskip

\begin{flushleft}
\textbf{Proof of Theorem \ref{thm:net_2sls}}
\end{flushleft}

Similar to \eqref{eq:pair_2sls_exp}, we can write
\begin{align*}
    \beta_s
    & = \frac{\sumi \bbE[(D_{1(i)} - \bbL(D_{1(i)} \mid D_i)) Y_i] }{\sumi \bbE[(D_{1(i)} - \bbL(D_{1(i)} \mid D_i)) M_i]} \\
    & = \frac{\sumi \bbE[(D_{1(i)} - p_1^\mathcal{J}) Y_i] }{\sumi \bbE[(D_{1(i)} - p_1^\mathcal{J}) M_i]} \\
    & = \frac{\sumi (\Cov(D_{1(i)}, Y_i \mid D_i = 1)p_1^\mathcal{I} + \Cov(D_{1(i)}, Y_i \mid D_i = 0)p_0^\mathcal{I})}{\sumi \bbE[(D_{1(i)} - p_1^\mathcal{J}) M_i]}.
\end{align*}
Moreover, $\Cov(D_{1(i)}, Y_i \mid D_i = d) =  \{\bbE[ Y_i \mid D_{1(i)} = 1, D_i = d] - \bbE[ Y_i \mid D_{1(i)} = 0, D_i = d]\}  p_1^\mathcal{J} p_0^\mathcal{J}$.

Let $\lambda_i(r, d_j) \coloneqq \bm{1}\{R_i(d_j) \ge r\}$ so that we can write 
\begin{align*}
    Y_i = \sum_{(d_i, d_j, r) \in \{0,1\}^2 \times \mathcal{R}_i} \bm{1}\{D_i = d_i, D_{1(i)} = d_j\} [\lambda_i(r, d_j) - \lambda_i(r + 1, d_j)] Y_i(d_i, r).
\end{align*}
Using this decomposition, we obtain
\begin{align*}
        \bbE[Y_i \mid D_{1(i)} = d_j, D_i = d_i]
        & = \sum_{r \in \mathcal{R}_i} \bbE[ \{\lambda_i(r, d_j) - \lambda_i(r + 1, d_j)\} Y_i(d_i, r)]
\end{align*}
from Assumption \ref{as:rct_net}.
Hence,
\begin{align*}
    & \bbE[ Y_i \mid D_{1(i)} = 1, D_i = d] - \bbE[ Y_i \mid D_{1(i)} = 0, D_i = d] \\
    & = \sum_{r \in \mathcal{R}_i} \bbE\left[ \{\lambda_i(r, 1) - \lambda_i(r, 0) - \lambda_i(r + 1, 1) + \lambda_i(r + 1, 0) \} Y_i(d, r) \right] \\
    & = \sum_{r = 1}^{n_i} \bbE\left[ \tau_i^1 (d, r) \{\lambda_i(r, 1) - \lambda_i(r, 0)\}  \right] \\
    & = \sum_{r = 1}^{n_i} \bbE\left[ \tau_i^1(d, r) \mid R_i(1) \ge r > R_i(0)  \right] \Pr( R_i(1) \ge r > R_i(0)),
\end{align*}
where the last equality follows from the fact that $\lambda_i(r, 1) - \lambda_i(r, 0)$ is either one or zero.
Thus, the numerator of $\beta_s$ is
\begin{align*}
\sumi \sum_{r = 1}^{n_i} \bbE\left[ \bar \tau_i^1(r) \mid R_i(1) \ge r > R_i(0)  \right] \Pr( R_i(1) \ge r > R_i(0)) p_1^\mathcal{J} p_0^\mathcal{J}.
\end{align*}
For the denominator, we note that
\begin{align*}
    \mu 
    = \frac{1}{n}\sumi \bbE[M_i]
    & = \frac{1}{n}\sumi \sum_{d \in \{0,1\}} \bbE[M_i \mid D_{1(i)} = d] p_d^\mathcal{J} \\
    & = \frac{1}{n}\sumi \sum_{d \in \{0,1\}} \bbE[M_i^d] p_d^\mathcal{J},
\end{align*}
we have
\begin{align*}
    \sumi \bbE[(D_{1(i)} - p_1^\mathcal{J}) M_i]
    & = \sumi \bbE[D_{1(i)} M_i] - n p_1^\mathcal{J} \mu \\
    & = \sumi \bbE[ M_i^1] p_1^\mathcal{J} - \sumi \bbE[M_i^1] (p_1^\mathcal{J})^2 - \sumi \bbE[M_i^0] p_1^\mathcal{J} p_0^\mathcal{J} \\
    & = \sumi \bbE[M_i^1 - M_i^0] p_1^\mathcal{J} p_0^\mathcal{J}.
\end{align*}
By combining these, the proof is complete.
\qed


\bigskip

\begin{flushleft}
\textbf{Proof of Theorem \ref{thm:net_sls}}
\end{flushleft}

(i) For the first-stage regression, the objective function can be rewritten as
\begin{align*}
    & \sumi\bbE[A_{1(i)}(M_i - a_0(1- D_i) - a_1 D_i - a_s D_{1(i)})^2]  \\
    & \quad = \sumi\bbE[A_{1(i)} (M_i - a_1 - a_s D_{1(i)})^2 D_i ] + \sumi\bbE[A_{1(i)} (M_i - a_0 - a_s D_{1(i)})^2 (1 - D_i) ].
\end{align*}
Solving the first-order conditions for $\gamma_{0,1}$ and $\gamma_{1,1}$ yields
\begin{align*}
    \gamma_{0,1} = \gamma_{1,1} = \mu^A /\pi^A - \gamma_{s,1} p_1^\mathcal{J},
\end{align*}
where $\mu^A \coloneqq n^{-1}\sumi\bbE[A_{1(i)} M_i]$, and $\pi^A \coloneqq n^{-1} \sumi \Pr(A_{1(i)} = 1)$.
Thus, we obtain
\begin{align*}
    0 
    & = \sumi \bbE[A_{1(i)} D_{1(i)} (M_i - \gamma_{0,1} (1 - D_i) - \gamma_{1,1} D_i - \gamma_{s,1} D_{1(i)})] \\
    & = \sumi \bbE[A_{1(i)} D_{1(i)} (M_i - \mu^A /\pi^A - \gamma_{s,1} [D_{1(i)} - p_1^\mathcal{J}])] \\
    & = \sumi \bbE[A_{1(i)} D_{1(i)} (M_i - \mu^A /\pi^A)] - n \pi^A p_1^\mathcal{J} p_0^\mathcal{J} \gamma_{s,1}
    \Longrightarrow \;\; \gamma_{s,1} = \frac{n^{-1}\sumi \bbE[A_{1(i)} D_{1(i)} (M_i - \mu^A /\pi^A)]}{ \pi^A p_1^\mathcal{J} p_0^\mathcal{J} }
\end{align*}
and
\begin{align*}
    \bbL_1(M_i \mid D_i, D_{1(i)}) = \mu^A /\pi^A + \frac{(D_{1(i)} - p_1^\mathcal{J}) n^{-1}\sumj \bbE[A_{1(j)} D_{1(j)} (M_j - \mu^A /\pi^A)]}{ \pi^A p_1^\mathcal{J} p_0^\mathcal{J} }.
\end{align*}

For the second-stage regression, the first-order condition for $\beta_1$ implies that 
\begin{align*}
    0  
    & = \sumi \bbE[A_{1(i)} (Y_i - \beta_1 - \beta_s \bbL_1(M_i \mid D_i, D_{1(i)})) D_i ]  \\
    \iff 0 & = \sumi \bbE[A_{1(i)} Y_i \mid D_i = 1] - \beta_1 \sumi \bbE[A_{1(i)}]- \beta_s \sumi \bbE[ \bbL_1(M_i \mid D_i, D_{1(i)}) \mid D_i = 1, A_{1(i)} = 1]  \bbE[A_{1(i)}] \\
    & = \sumi  \bbE[A_{1(i)} Y_i \mid D_i = 1] - n \beta_1 \pi^A - n \beta_s \mu^A,
\end{align*}
and thus $\beta_1  = n^{-1} \sumi \bbE[A_{1(i)} Y_i \mid D_i = 1]/\pi^A  - \beta_s \mu^A /\pi^A$ and $\beta_0  = n^{-1} \sumi \bbE[A_{1(i)} Y_i \mid D_i = 0]/\pi^A - \beta_s \mu^A /\pi^A$.
Noting that
\begin{align*}
    \bbE[A_{1(i)} Y_i \mid D_i = d] 
    & = \sum_{(r, \vec{a}) \in \mathcal{R}_i \times \{1, \mathcal{A}_{i,-1}\}} \bbE[Y_i \mid R_i = r, \bm{A}_{\mathcal{P}_i} = \vec{a}, D_i = d] \Pr(R_i = r, \bm{A}_{\mathcal{P}_i} = \vec{a} \mid D_i = d) \\
    & = \sum_{(r, \vec{a}) \in \mathcal{R}_i \times \{1, \mathcal{A}_{i,-1}\}} \bbE[Y_i(d,r) \mid \bm{A}_{\mathcal{P}_i} = \vec{a}] \Pr(R_i = r, \bm{A}_{\mathcal{P}_i} = \vec{a}),
\end{align*}
we have
\begin{align*}
    \beta_d = \frac{n^{-1}\sumi \sum_{(r, \vec{a}) \in \mathcal{R}_i \times \{1, \mathcal{A}_{i,-1}\}} \bbE[\delta_i(r) \mid \bm{A}_{\mathcal{P}_i} = \vec{a}] \Pr(R_i = r, \bm{A}_{\mathcal{P}_i} = \vec{a})}{\pi^A}.
\end{align*}

\bigskip

(ii) Let $\bbL_1(D_{1(i)} \mid D_i)$ be the weighted linear projection of $D_{1(i)}$ onto $D_i$ with weight $A_{1(i)}$.
As $D_{1(i)}$ is independent of $(D_i, A_{1(i)})$, we obtain $\bbL_1(D_{1(i)} \mid D_i) = p_1^\mathcal{J}$.
Then, similar to \eqref{eq:pair_2sls_exp}, we can write
\begin{align*}
    \beta_s
    & = \frac{\sumi \bbE[A_{1(i)} (D_{1(i)} - \bbL_1(D_{1(i)} \mid D_i)) Y_i] }{\sumi \bbE[A_{1(i)} (D_{1(i)} - \bbL_1(D_{1(i)} \mid D_i)) M_i]} \\
    & = \frac{\sumi \bbE[A_{1(i)} (D_{1(i)} - p_1^\mathcal{J}) Y_i] }{\sumi \bbE[A_{1(i)} (D_{1(i)} - p_1^\mathcal{J}) M_i]} \\
    & = \frac{\sumi (\Cov(D_{1(i)}, Y_i \mid A_{1(i)} = 1, D_i = 1) \bbE[A_{1(i)}] p_1^\mathcal{I} + \Cov(D_{1(i)}, Y_i \mid A_{1(i)} = 1, D_i = 0) \bbE[A_{1(i)}] p_0^\mathcal{I})}{\sumi \bbE[A_{1(i)} (D_{1(i)} - p_1^\mathcal{J}) M_i]}.
\end{align*}
It is clear that $\Cov(D_{1(i)}, Y_i \mid A_{1(i)} = 1, D_i = d) =  \{\bbE[ Y_i \mid D_{1(i)} = 1, A_{1(i)} = 1, D_i = d] - \bbE[ Y_i \mid D_{1(i)} = 0, A_{1(i)} = 1, D_i = d]\}  p_1^\mathcal{J} p_0^\mathcal{J}$.

By using the same decomposition as in the proof of Theorem \ref{thm:net_2sls}, we can write
\begin{align*}
        \bbE[Y_i \mid D_{1(i)} = d_j, A_{1(i)} = 1, D_i = d_i]
        & = \sum_{r \in \mathcal{R}_i} \bbE[ \{\lambda_i(r, d_j) - \lambda_i(r + 1, d_j)\} Y_i(d_i, r) \mid A_{1(i)} = 1].
\end{align*}
Hence,
\begin{align*}
    & \bbE[ Y_i \mid D_{1(i)} = 1, A_{1(i)} = 1, D_i = d] - \bbE[ Y_i \mid D_{1(i)} = 0, A_{1(i)} = 1, D_i = d] \\
    & = \sum_{r \in \mathcal{R}_i} \bbE\left[ \{\lambda_i(r, 1) - \lambda_i(r, 0) - \lambda_i(r + 1, 1) + \lambda_i(r + 1, 0) \} Y_i(d, r) \mid A_{1(i)} = 1 \right] \\
    & = \sum_{r = 1}^{n_i} \bbE\left[ \tau_i^1 (d, r) \{\lambda_i(r, 1) - \lambda_i(r, 0)\}  \mid A_{1(i)} = 1 \right] \\
    & = \sum_{r = 1}^{n_i} \bbE\left[ \tau_i^1(d, r) \mid R_i(1) \ge r > R_i(0)  \right] \Pr( R_i(1) \ge r > R_i(0) \mid A_{1(i)} = 1).
\end{align*}
Thus, the numerator of $\beta_s$ is
\begin{align*}
    \sumi \sum_{r = 1}^{n_i} \bbE\left[ \bar \tau_i^1(r) \mid R_i(1) \ge r > R_i(0)  \right] \Pr( R_i(1) \ge r > R_i(0)) p_1^\mathcal{J} p_0^\mathcal{J}.
\end{align*}
For the denominator, noting that
\begin{align*}
    \mu^A 
    = \frac{1}{n} \sumi \bbE[M_i \mid A_{1(i)} = 1] \bbE[A_{1(i)}]
    & = \frac{1}{n} \sumi \sum_{d \in \{0,1\}} \bbE[M_i \mid D_{1(i)} = d, A_{1(i)} = 1] \bbE[A_{1(i)}] p_d^\mathcal{J} \\
    & = \frac{1}{n} \sumi \sum_{d \in \{0,1\}} \bbE[M_i^d \mid A_{1(i)} = 1] \bbE[A_{1(i)}] p_d^\mathcal{J},
\end{align*}
we have
\begin{align*}
    \sumi \bbE[A_{1(i)} (D_{1(i)} - p_1^\mathcal{J}) M_i] 
    & = \sumi \bbE[ M_i \mid D_{1(i)} = 1, A_{1(i)} = 1] \bbE[A_{1(i)}] p_1^\mathcal{J} - n p_1^\mathcal{J} \mu^A \\
    & = \sumi \bbE[ M_i^1 \mid A_{1(i)} = 1] \bbE[A_{1(i)}] p_1^\mathcal{J} - \sumi \bbE[M_i^1 \mid A_{1(i)} = 1] \bbE[A_{1(i)}] (p_1^\mathcal{J})^2 \\
    & \quad - \sumi  \bbE[M_i^0 \mid A_{1(i)} = 1] \bbE[A_{1(i)}] p_0^\mathcal{J} p_1^\mathcal{J} \\
    & = \sumi \bbE[M_i^1 - M_i^0 \mid A_{1(i)} = 1] \Pr( A_{1(i)} = 1) p_1^\mathcal{J} p_0^\mathcal{J}.    
\end{align*}
Furthermore, 
\begin{align*}
   \bbE[M_i^1 - M_i^0] 
   & = \bbE[M_i^1 - M_i^0 \mid A_{1(i)} = 1] \Pr( A_{1(i)} = 1) + \bbE[M_i^1 - M_i^0 \mid A_{1(i)} = 0] \Pr( A_{1(i)} = 0) \\
   & = \bbE[M_i^1 - M_i^0 \mid A_{1(i)} = 1] \Pr( A_{1(i)} = 1).
\end{align*}
By combining these, the proof is complete.

\qed

\bigskip


\begin{flushleft}
\textbf{Derivation of \eqref{eq:clt_net_2sls} and \eqref{eq:clt_net_sls}}
\end{flushleft}

Observe that
\begin{align*}
    \sqrt{n}(\hat \beta_s^{2sls} - \beta_s^{2sls})
    & = \sqrt{n}( \tilde{\bm{D}}_{1,n}^\top (\bm{D}_{c,n} \beta_c^{2sls} + \bm{M}_n \beta_s^{2sls} + \bm{\varepsilon}_n) / \tilde{\bm{D}}_{1,n}^\top \bm{M}_n - \beta_s^{2sls})\\
    & = ( \tilde{\bm{D}}_{1,n}^\top \bm{\varepsilon}_n / \sqrt{n}) / ( \tilde{\bm{D}}_{1,n}^\top \bm{M}_n / n),
\end{align*}
where $\bm{\varepsilon}_n = (\varepsilon_1, \ldots, \varepsilon_n)^\top$, and $\beta_c^{2sls} = (\beta_0^{2sls}, \beta_d^{2sls})^\top$, and that
\begin{align*}
    \sqrt{n}(\hat \beta_s^{wls} - \beta_s^{wls})
    & = \sqrt{n}( \tilde{\bm{D}}_{1,n,A}^\top (\bm{D}_{c,n} \beta_c^{wls} + \bm{M}_n \beta_s^{wls} + \bm{\epsilon}_{1,n}) / \tilde{\bm{D}}_{1,n,A}^\top \bm{M}_n - \beta_s^{wls})\\
    & = ( \tilde{\bm{D}}_{1,n,A}^\top \bm{\epsilon}_{1,n} / \sqrt{n}) / ( \tilde{\bm{D}}_{1,n,A}^\top \bm{M}_n /n ),
\end{align*}
where $\bm{\epsilon}_{1,n} = (\epsilon_{1,1}, \ldots, \epsilon_{1,n})^\top$, and $\beta_c^{wls} = (\beta_0^{wls}, \beta_d^{wls})^\top$.

Below, we discuss only the derivation of the asymptotic distribution of the WLS estimator because the 2SLS estimator is analogous.
Throughout, we assume the existence of the third moment of $\epsilon_{1,i}$.
Under Assumptions \ref{as:rct_net} and \ref{as:id_net}, we can easily show that  $\bm{D}_{c,n}^\top \mathbb{I}_{n, A} \bm{D}_{c,n}/n = \pi^A \left(\begin{array}{cc} 1 & p_1^\mathcal{I} \\ p_1^\mathcal{I} & p_1^\mathcal{I} \end{array}\right) + O_P(n^{-1/2})$ and $\bm{D}_{c,n}^\top \mathbb{I}_{n, A} \bm{D}_{J,n}/n = \pi^A \left(\begin{array}{c} p_1^\mathcal{J} \\ p_1^\mathcal{I} p_1^\mathcal{J} \end{array}\right) + O_P(n^{-1/2})$, leading to $(\bm{D}_{c,n}^\top \mathbb{I}_{n, A} \bm{D}_{c,n}/n)^{-1} \bm{D}_{c,n}^\top \mathbb{I}_{n, A} \bm{D}_{J,n}/n = (p_1^\mathcal{J}, 0)^\top + o_P(1)$.
For the denominator on the right-hand side, we observe that
\begin{align*}
        \tilde{\bm{D}}_{1,n,A}^\top \bm{M}_n/n
        & = \bm{M}_n^\top \mathbb{I}_{n, A} \bm{D}_{1,n}/n - \bm{M}_n^\top \mathbb{I}_{n, A} \bm{D}_{c,n}/n  (\bm{D}_{c,n}^\top \mathbb{I}_{n, A} \bm{D}_{c,n}/n)^{-1} \bm{D}_{c,n}^\top \mathbb{I}_{n, A} \bm{D}_{1,n}/n \\
        & = \frac{1}{n}\sumi M_i A_{1(i)} D_{1(i)}  - \frac{1}{n}\sumi M_i A_{1(i)} D_{c,i}^\top (\bm{D}_{c,n}^\top \mathbb{I}_{n, A} \bm{D}_{c,n}/n)^{-1} \bm{D}_{c,n}^\top \mathbb{I}_{n, A} \bm{D}_{1,n}/n \\
        & = \frac{1}{n}\sumi \bbE[M_i A_{1(i)} (D_{1(i)} - p_1^\mathcal{J})] + o_P(1)\\
        & = \frac{1}{n}\sumi \bbE[M_i^1 - M_i^0] p_1^\mathcal{J} p_0^\mathcal{J} + o_P(1).
\end{align*}
Moreover, noting that $||n^{-1/2} \sumi D_{c,i}^\top \epsilon_i|| = O_P(1)$ under $\bbE[\epsilon_i] = \bbE[D_i \epsilon_i]$ = 0, we have
\begin{align*}
        \tilde{\bm{D}}_{1,n,A}^\top \bm{\epsilon}_{1,}n /\sqrt{n}
        & = \frac{1}{\sqrt{n}}\sumi D_{1(i)} \epsilon_i - (\bm{D}_{1,n}^\top \mathbb{I}_{n, A} \bm{D}_{c,n}/n) (\bm{D}_{c,n}^\top \mathbb{I}_{n, A} \bm{D}_{c,n}/n)^{-1} \left( \frac{1}{\sqrt{n}} \sumi D_{c,i}^\top \epsilon_i \right) \\
        & = \frac{1}{\sqrt{n}} \sumi (D_{1(i)} - p_1^\mathcal{J}) A_{1(i)}\epsilon_{1,i} + o_P(1).
\end{align*}
Furthermore, from Assumptions \ref{as:id_net} and $\bbE[D_{1(i)} \epsilon_i] = 0$,
\begin{align*}
    \Var\left(\frac{1}{\sqrt{n}} \sumi (D_{1(i)} - p_1^\mathcal{J}) A_{1(i)}\epsilon_{1,i} \right) 
    & = \frac{1}{n} \sumi \bbE\left[(D_{1(i)} - p_1^\mathcal{J})^2 A_{1(i)} \epsilon_{1,i}^2 \right] \\
    & = (p_0^\mathcal{J})^2 p_1^\mathcal{J} \cdot \frac{1}{n} \sumi \sigma^2_{\epsilon, 1, i}(1) \Pr(A_{1(i)} = 1)  + p_0^\mathcal{J} (p_1^\mathcal{J})^2 \cdot \frac{1}{n} \sumi \sigma^2_{\epsilon, 1, i}(0) \Pr(A_{1(i)} = 1).
\end{align*}
Finally,  by Lyapunov's central limit theorem and Slutsky's theorem, we obtain
\begin{align*}
    \sqrt{n}(\hat \beta_s^{wls} - \beta_s^{wls})
    \overset{d}{\to} N \left(0, \lim_{n\to\infty} \frac{p_0^\mathcal{J} \frac{1}{n}\sumi \sigma^2_{\epsilon, 1, i}(1) \Pr(A_{1(i)} = 1) + p_1^\mathcal{J}  \frac{1}{n}\sumi \sigma^2_{\epsilon, 1, i}(0) \Pr(A_{1(i)} = 1)}{\left(\frac{1}{n}\sumi \bbE[M_i^1 - M_i^0] \right)^2 p_1^\mathcal{J} p_0^\mathcal{J}}\right).
\end{align*}

\qed

\subsection{Derivation of the asymptotic normality results in Section \ref{sec:inference}}\label{app:proof_sec4}

We discuss only the results of the WLS estimator.
In addition, to simplify the notation, we suppress condition $i \in \mathcal{I}'$ in expectations when there is no confusion.
The population WLS estimand is then defined as
\begin{align*}
    (\beta_0, \beta_d, \beta_s) = \argmin_{b_0, b_d, b_s} \bbE[A_{1(i)}(Y_i - b_0 - b_d D_i - b_s \mathbb{L}_1(M_i \mid D_i, D_{1(i)}))^2],
\end{align*}
where $\mathbb{L}_1(M_i \mid D_i, D_{1(i)}) \coloneqq \gamma_{0,1} + \gamma_{d,1} D_i + \gamma_{s,1} D_{1(i)}$, and
\begin{align*}
    (\gamma_{0,1}, \gamma_{d,1}, \gamma_{s,1}) = \argmin_{a_0, a_d, a_s}  \bbE[A_{1(i)}(M_i - a_0 - a_d D_i - a_s D_{1(i)})^2].
\end{align*}
From the same discussion as in the proof of Theorem \ref{thm:net_sls}, we can see that
\begin{align*}
    \gamma_{0,1} 
    & = \gamma_{1,1} = m^A - \gamma_{s,1} p_1^\mathcal{J} \\
    \gamma_{s,1}
    & = \frac{\bbE[ A_{1(i)}D_{1(i)} (M_i - m^A)] }{q_1^A p_1^\mathcal{J} p_0^\mathcal{J}} \\
    \mathbb{L}_1(M_i \mid D_i, D_{1(i)})
    & = m^A + \frac{(D_{1(i)} - p_1^\mathcal{J}) \bbE[ A_{1(i)}D_{1(i)} (M_i - m^A)] }{q_1^A p_1^\mathcal{J} p_0^\mathcal{J}}
\end{align*}
where $q_1^A \coloneqq \Pr(A_{1(i)} = 1 \mid i \in \mathcal{I}')$, and $m^A \coloneqq \bbE[M_i \mid A_{1(i)} = 1, i \in \mathcal{I}']$.
Further,
\begin{align*}
    \beta_1
    & = \bbE[Y_i \mid D_i = 1, A_{1(i)} = 1] - \beta_s m^A \\
    \beta_0
    & = \bbE[Y_i \mid D_i = 0, A_{1(i)} = 1] - \beta_s m^A. 
\end{align*}
Thus, we can write
\begin{align*}
    \epsilon_i 
    & = A_{1(i)} \epsilon_{1,i} \\
    & = A_{1(i)} (Y_i - \beta_0 - D_i \beta_d - M_i \beta_s )\\
    & = A_{1(i)} D_i \sum_{r \in \mathcal{R}_i}\bm{1}\{R_i = r\} Y_i(1, r) -  A_{1(i)}D_i \bbE[Y_i \mid D_i = 1, A_{1(i)} = 1] \\
    & \quad + A_{1(i)} (1 - D_i) \sum_{r \in \mathcal{R}_i}\bm{1}\{R_i = r\} Y_i(0, r) - A_{1(i)} (1 - D_i) \bbE[Y_i \mid D_i = 0, A_{1(i)} = 1] - \beta_s A_{1(i)} ( M_i - m^A) \\
    & = A_{1(i)} D_i \sum_{r \in \mathcal{R}_i} \left\{ \bm{1}\{R_i = r\} Y_i(1, r) -  \bbE[Y_i(1, r) \mid A_{1(i)} = 1, R_i = r] \Pr(R_i = r \mid A_{1(i)} = 1) \right\}\\
    & \quad + A_{1(i)} (1 - D_i) \sum_{r \in \mathcal{R}_i} \left\{ \bm{1}\{R_i = r\} Y_i(0, r) - \bbE[Y_i(0, r) \mid A_{1(i)} = 1, R_i = r] \Pr(R_i = r \mid A_{1(i)} = 1) \right\} \\
    & \quad - \beta_s A_{1(i)} ( M_i - m^A).
\end{align*}
From the last line, it is clear that $\bbE[\epsilon_i] = 0$.
Similarly, since
\begin{align*}
    D_i \epsilon_i 
    & = A_{1(i)} D_i \sum_{r \in \mathcal{R}_i} \left\{ \bm{1}\{R_i = r\} Y_i(1, r) -  \bbE[Y_i(1, r) \mid A_{1(i)} = 1, R_i = r] \Pr(R_i = r \mid A_{1(i)} = 1) \right\}\\
    & \quad - \beta_s A_{1(i)} D_i ( M_i - m^A),
\end{align*}
$\bbE[D_i \epsilon_i] = 0$ can be easily confirmed as well.
Finally, $\bbE[D_{1(i)} \epsilon_i] = 0$ is straightforward from plugging the right-hand side of $Y_i = \beta_0 + \beta_d D_i + \beta_s M_i + \epsilon_{1,i}$ into $\beta_s = \frac{\bbE[A_{1(i)} (D_{1(i)} - p_1^\mathcal{J}) Y_i \mid i \in \mathcal{I}'] }{\bbE[A_{1(i)} (D_{1(i)} - p_1^\mathcal{J}) M_i \mid i \in \mathcal{I}']}$.

The remainder of the discussion is identical to the derivation of \eqref{eq:clt_net_sls}.
\qed


\section{Replication R code}\label{app:Rcode}

\begin{lstlisting}[basicstyle=\ttfamily\footnotesize, frame=single]
    library(AER)
    set.seed(2023)
    N <- 1000
    Di <- ifelse(runif(N) < 0.5, 1, 0)
    Dj <- ifelse(runif(N) < 0.5, 1, 0)
    Ui <- runif(N, -1, 1)
    Uj <- runif(N, -1, 1)
    A  <- ifelse(plogis(Ui + Uj) > 0.5, 1, 0)
    S  <- A*Dj
    Y  <- rnorm(N, mean = 1, sd = 1) + Ui
   
    # OLS 
    summary(lm(Y ~ Di + S))
   
    # 2SLS
    summary(ivreg(Y ~ Di + S | Di + Dj))
   
    # WLS 
    summary(lm(Y ~ Di + S, weight = A))
\end{lstlisting}


\section{Monte Carlo experiments}\label{app:MC}

In this appendix, we investigate the finite sample performance of the proposed method.
We consider the following data-generating process (DGP) for two sample sizes $n \in \{400, 1600\}$:
\begin{align*}
    Y_i = \sum_{(d,r) \in \{0,1\} \times \mathcal{R}_i} \bm{1}\{D_i = d, R_i = r\} Y_i(d,r), \;\; \text{where} \;\; Y_i(d,r) = \beta_{0,i} + \beta_{d,i} d + \beta_{s,i} r + U_i,
\end{align*}
$n_i \in \{1,3,6,9\}$, which is fixed throughout the simulations such that $n_i = 1$ for the first quarter of the sample, $n_i = 3$ for the second quarter, and so forth; $D_i, D_{j(i)} \sim \text{Bernoulli}(0.5)$, $A_{j(i)} = \bm{1}\{e^{(U_i + U_{j(i)})}/[1 + e^{(U_i + U_{j(i)})}] > c\}$ with $c \in \{0.3, 0.6\}$ and $U_i, U_{j(i)} \sim N(0,1)$, and
\begin{align*}
    & \beta_{0,i} \sim 3 \cdot \bm{1}\{n_i \in \{1,3\} \} + \bm{1}\{n_i \in \{6,9\} \} + N(0,1) \\
    & \beta_{d,i} \sim N(0,4) \\
    & \beta_{s,i} \sim h \cdot (\text{Uniform}(0,2) + \bm{1}\{A_{1(i)} = 1 \}\text{Uniform}(0,1))
\end{align*}
with $h \in \{0,1\}$.
With this DGP, since $\bbE[\bar \tau_i^1(r) \mid R_i(1) \ge r > R_i(0)] = h \cdot 1.5$ uniformly in $i$ and $r$, the true LATE parameter is $\beta_s^{2sls} = \beta_s^{wls} = h \cdot 1.5$.

First, we examine the performance of the 2SLS and WLS estimators.
For comparison, we also perform an OLS estimation when $h = 0$ (i.e., no spillover effects).
The number of Monte Carlo repetitions for each scenario is set to 1000, and the estimators are evaluated in terms of bias and RMSE (root mean squared error).
In addition, based on the asymptotic normality results in \eqref{eq:clt_net_2sls} and \eqref{eq:clt_net_sls} with the sample analog estimation of asymptotic variances, we compute the 95\% confidence intervals and report their simulated coverage rates.
The results are summarized in Table \ref{table:MCest}.

Our main findings from Table \ref{table:MCest} are as follows.
As expected, the 2SLS and WLS estimators work reasonably well in all settings. However, the OLS estimator, which does not account for network endogeneity, is severely biased (the true value for the OLS estimator is set to zero).
In particular, when the probability of network connection is low, the WLS estimator clearly outperforms the 2SLS estimator in terms of RMSE, which is consistent with our theory.
Except for some situations with small sample sizes, the coverage rates of the confidence intervals are close to the nominal 95\% coverage.
Note that in this experimental design, the population residuals have non-identical means (see the definition of $\beta_{0,i}$), in which the asymptotic normal approximation may not be precise in general.
We presume that these results are specific to our chosen DGP; however, a more formal investigation is required.

\begin{table}[!ht]
    \begin{center}
    \caption{Performance of 2SLS and WLS}\label{table:MCest}
    \begin{tabular}{ccc|cc|ccc|ccc}
    \hline\hline
         & & & \multicolumn{2}{c|}{OLS} & \multicolumn{3}{c|}{2SLS} & \multicolumn{3}{c}{WLS} \\ 
        $n$ & $h$ & $c$ & Bias & RMSE & Bias & RMSE & CovRatio & Bias & RMSE & CovRatio \\ \hline
        400 & 0 & 0.3 & -0.1623 & 0.1623 & 0.0114 & 0.2175 & 0.930 & 0.0119 & 0.1707 & 0.953 \\ 
        & & 0.6 & 0.0675 & 0.0754 & -0.0679 & 0.5462 & 0.930 & 0.0290 & 0.2455 & 0.925 \\
        1600 & 0 & 0.3 & -0.1601 & 0.1601 & 0.0060 & 0.0991 & 0.958 & 0.0090 & 0.0786 & 0.954 \\ 
        & & 0.6 & 0.0701 & 0.0702 & 0.0068 & 0.1888 & 0.955 & 0.0109 & 0.1074 & 0.956 \\
        400 & 1 & 0.3 & & & 0.0005 & 0.2924 & 0.944 & 0.0005 & 0.2324 & 0.946 \\
        & & 0.6 & & & -0.0722 & 0.6138 & 0.930 & 0.0135 & 0.3110 & 0.925 \\
        1600 & 1 & 0.3 & & & 0.0012 & 0.1369 & 0.959 & 0.0047 & 0.1062 & 0.951 \\
        & & 0.6 & & & 0.0035 & 0.2194 & 0.957 & 0.0078 & 0.1333 & 0.952 \\ \hline
    \end{tabular}
    \end{center}
\end{table}

Next, we study the performance of the randomization test proposed in Subsection \ref{subsec:frt}.
For the choice of the test statistics, we consider the following four alternatives: 2SLS, WLS, ITT, and ITTC.
In this experiment, except that $c = 0.5$ is fixed, the remaining DGP settings are the same as above.
The number of simulations used to estimate the $p$-value is set to $B = 500$.
Table \ref{table:MCtest} reports the simulation results for the rejection frequency of these four statistics over 1000 Monte Carlo repetitions at the 10\%, 5\%, and 1\% significance levels.
The results show that our randomization test performs satisfactorily in all cases with reasonably accurate size control.
The WLS statistic showed the best performance in terms of test power.
The ITT statistic has relatively low power, which is a legitimate result considering that the definition of this statistic ignores network connectivity.

\begin{table}[!ht]
    \begin{center}
    \caption{Rejection frequency}\label{table:MCtest}
    \begin{tabular}{cc|cccccc}
    \hline\hline
         &  & \multicolumn{3}{c}{2SLS}  & \multicolumn{3}{c}{WLS}  \\
        $n$ & $h$ & 10\% & 5\% & 1\% & 10\% & 5\% & 1\% \\
        \hline
        400 & 0 & 0.105  & 0.061  & 0.013  & 0.115  & 0.059  & 0.011  \\
        1600 &  & 0.105  & 0.045  & 0.010  & 0.099  & 0.040  & 0.010  \\
        400 & 1 & 0.946  & 0.913  & 0.800  & 0.999  & 0.996  & 0.983  \\
        1600 &  & 1.000  & 1.000  & 1.000  & 1.000  & 1.000  & 1.000  \\
        \hline\hline
         &  & \multicolumn{3}{c}{ITT}  & \multicolumn{3}{c}{ITTC}  \\
        $n$ & $h$ & 10\% & 5\% & 1\% & 10\% & 5\% & 1\% \\
         \hline
        400 & 0 & 0.109  & 0.056  & 0.013  & 0.113  & 0.058  & 0.008  \\
        1600 &  & 0.104  & 0.048  & 0.006  & 0.085  & 0.041  & 0.008  \\
        400 & 1 & 0.867  & 0.798  & 0.549  & 0.990  & 0.976  & 0.891  \\
        1600 &  & 1.000  & 1.000  & 0.997  & 1.000  & 1.000  & 1.000  \\
         \hline
    \end{tabular}
    \end{center}
\end{table}

\end{document}